\pdfoutput=1

\documentclass[11pt,twoside,a4paper,cmspaper,final,collab]{cms-tdr}
\def\svnVersion{220873}\def\cmsCernNo{2013-133}\def\cmsCernDate{\today}\def\cmsMessage{Published in the Journal of High Energy Physics as \href{http://dx.doi.org/10.1007/JHEP10(2013)167}{\doi{10.1007/JHEP10(2013)167.}}}

\begin{document}\cmsNoteHeader{TOP-11-020}

\hyphenation{had-ron-i-za-tion}
\hyphenation{cal-or-i-me-ter}
\hyphenation{de-vices}

\RCS$Revision: 212427 $
\RCS$HeadURL: svn+ssh://svn.cern.ch/reps/tdr2/papers/TOP-11-020/trunk/TOP-11-020.tex $
\RCS$Id: TOP-11-020.tex 212427 2013-10-18 14:38:48Z mara $
\renewcommand\Im{\operatorname{\mathrm{Im}}}
\renewcommand\Re{\operatorname{\mathrm{Re}}}
\providecommand{\MT}{\ensuremath{M_\mathrm{T}}\xspace}

\cmsNoteHeader{TOP-11-020}

\title{Measurement of the W-boson helicity in top-quark decays from \ttbar production in lepton+jets events in pp collisions at $\sqrt{s}=7$\TeV}

\date{\today}

\abstract{
The W-boson helicity fractions in
top-quark decays are measured with \ttbar events in the lepton+jets final state,
using proton-proton collisions at a centre-of-mass energy of 7\TeV,
collected in 2011 with the CMS detector at the LHC. The data sample corresponds to an
integrated luminosity of  5.0\fbinv. The measured fractions of longitudinal, left-, and right-handed helicity are $F_0=0.682 \pm 0.030\stat\pm 0.033\syst$, $F_L=0.310 \pm 0.022\stat\pm 0.022\syst$,
and  $F_R= 0.008 \pm 0.012\stat\pm 0.014\syst$,
consistent with the standard model predictions.  The measured fractions are
used to probe the existence of anomalous $\PW\cPqt\cPqb$ couplings. Exclusion limits on the real components of the anomalous couplings $g_L,\ g_R$ are also derived.
}

\hypersetup{%
pdfauthor={CMS Collaboration},
pdftitle={Measurement of the W-boson helicity in top-quark decays from ttbar production in lepton+jets events in pp collisions at sqrt(s)=7 TeV},
pdfsubject={CMS},
pdfkeywords={CMS, physics, W boson, helicity, top decays}}

\newcommand{\costh}{$\cos \theta^{*}$}
\newcommand{\costhhad}{\ensuremath{\cos^\text{had} \theta^{*}}}

\maketitle 

\section{Introduction}
\label{sec:intro}

Following the discovery of the top quark in proton-antiproton collisions at the Tevatron collider~\cite{discovCDF,discovD0},
measurements of W-boson helicity fractions in top-quark decays
have
been an important subject of
investigation, because of their relationship to the V$-$A structure of the
weak charged current and their sensitivity to physics beyond the standard model (SM).
With the large samples of \ttbar events produced in proton-proton
collisions at the Large Hadron Collider (LHC), the W-boson helicity
fraction measurements can be  improved considerably, enhancing the search for
anomalous Wtb couplings, \ie those that do not arise from the SM.
Previous measurements of  W-boson helicity fractions in top-quark decays have been performed by the
CDF, \DZERO~\cite{TevComb}, and ATLAS~\cite{AtlasPaper} Collaborations.

The helicity fractions of the W boson produced in a $\cPqt\to\PW\cPqb$ decay are defined as the
partial rate for a given helicity state
divided by the total decay rate: $F_{L,R,0}\equiv \Gamma_{L,R,0}/\Gamma$, where $F_L$, $F_R$, and $F_0$ are the left-handed, right-handed, and
longitudinal helicity fractions, respectively.
For SM couplings and unpolarised top-quark production, the helicity fractions are approximately
70\% longitudinal and 30\%
left-handed. At leading order (LO) and in the limit $m_\cPqb=0$
(where $m_\cPqb$ is the b-quark mass), the right-handed helicity fraction
is zero due to helicity suppression.
For  finite $m_\cPqb$, the helicity fractions are
\cite{WHelFracLO} :
\begin{align}
F_0 &= \frac{ (1-y^2)^2 - x^2(1+y^2)}{(1-y^2)^2 +x^2(1-2x^2+y^2)},  \\
F_L &= \frac{ x^2(1-x^2+y^2+\sqrt\lambda)}{(1-y^2)^2+x^2(1-2x^2+y^2)}, \\
F_R &= \frac{x^2(1-x^2+y^2-\sqrt\lambda)}{(1-y^2)^2+x^2(1-2x^2+y^2) },
\label{masseq}
\end{align}
where $x=M_\PW/m_\cPqt$, $y=m_\cPqb/m_\cPqt$,
$\lambda=1+x^4+y^4-2x^2y^2-2x^2-2y^2$,  and $M_\PW$, $m_\cPqt$ are
the masses of the W boson and top quark, respectively.
These fractions are minimally modified by higher-order corrections.
Recent next-to-next-to-leading-order (NNLO) calculations~\cite{Czarnecki:2010gb}
predict $F_0=0.687 \pm 0.005$,
$F_L=0.311 \pm 0.005$, and $F_R=0.0017 \pm
  0.0001$  for a top-quark mass of
$m_\cPqt=172.8\pm 1.3$\GeVcc.

Experimentally, the W-boson helicity components can be extracted through the study of angular
distributions of top-quark decay products in \ttbar final states.
The helicity angle $\theta^*$  is defined as the angle between the W-boson momentum in the top-quark rest frame and the momentum of the down-type decay fermion in the rest frame of the W boson.
The distribution of the cosine of the helicity angle has a dependence on the helicity fractions given by:
\begin{equation}
\frac{1}{\Gamma}\frac{\rd\Gamma}{\rd\cos{\theta^*}} =
\frac{3}{8} F_L  \left(1-\cos{\theta^*}\right)^2  +
\frac{3}{4} F_0 (\sin{\theta^*} )^2  +
\frac{3}{8} F_R \left(1+\cos{\theta^*}\right)^2 .
\label{eq:costhetastar}
\end{equation}

Deviations of the measured helicity fractions from the SM predictions can be
interpreted in terms of anomalous Wtb couplings~\cite{eqSaav1,eqSaav2}, using the most general dimension-six Lagrangian:

\begin{equation}
\mathcal{L}_{\PW\cPqt\cPqb}  =  - \frac{g}{\sqrt 2} \bar{b}  \gamma^{\mu} ( V_L P_L + V_R P_R ) t W_\mu^-
 - \frac{g}{\sqrt 2} \bar {b}  \frac{i \sigma^{\mu \nu} q_\nu}{M_\PW}  \left( g_L P_L + g_R P_R \right) t\; W_\mu^- + \text{h.c.} ,
\label{eq:Wtb0}
\end{equation}
\noindent
where $V_L,\ V_R,\ g_L,\ g_R$ are dimensionless complex constants, $q=p_\cPqt-p_\cPqb$, where $p_\cPqt$ ($p_\cPqb$) is
the four-momentum of the top quark (b quark),  $P_L\ (P_R)$ are the left (right) projector operators, and h.c.\ denotes the Hermitian conjugate.
Hermiticity conditions on the possible dimension-six Lagrangians also impose
$\Im(V_L)=0$~\cite{eqSaav2}.
In the SM and at tree level, $V_L = V_{\cPqt\cPqb}$,  where $V_{\cPqt\cPqb}$ is the Cabibbo--Kobayashi--Maskawa matrix element $V_{\cPqt\cPqb}\simeq 1$,  and $V_R=g_L=g_R=0$.
The relation between helicity fractions and anomalous couplings, including
dependencies on the b-quark mass, are given in ref.~\cite{SaavedraBernabeu}.

We report on a study of the W-boson helicity fractions in top-quark decays using
a sample of \ttbar events
where one of the top
quarks decays semileptonically (\eg $\cPqt\to\PW^+\cPqb\to \ell^+\nu_\ell\cPqb$, where $\ell$ is either an electron or a muon)
and the other decays hadronically
(\eg $\cPaqt\to\PWm\cPaqb\to
\Pq\Paq^\prime\cPaqb$).
A kinematic fit is used to determine the best association of
b jets, other jets, and lepton candidates to the top quark and
antiquark decay hypotheses,
interpreting the
measured momentum imbalance as  due to the presence of a
neutrino.  In this kinematic fit, top-quark and W-boson mass
constraints are employed to improve the resolution of the measured jet
and lepton energies, resulting in an improved
reconstruction of the W-boson rest frame and the helicity angles
in the weak decays of top quarks.
In this article, the cosine of the helicity angles in the semileptonic and hadronic
top-quark decays will be referred to as \costh and \costhhad, respectively. In the leptonic
branch, the down-type fermion corresponds to the charged lepton.
In the hadronic branch, since the down-type quark is not experimentally identified,
only the absolute value of \costhhad~is used, providing information on the relative
proportion between longitudinal ($F_0$) and total transverse ($F_L+F_R = 1-F_0$)
fractions.
The resulting
helicity angle distributions
are fitted to measure the W-boson helicity fractions and to determine possible
anomalous Wtb couplings.

\section{The CMS detector}
\label{sec:detector}

The central feature of the Compact Muon Solenoid (CMS) detector is
a superconducting solenoid,
13\unit{m} in length and 6\unit{m} in   diameter, which provides a uniform axial magnetic field of 3.8\unit{T}.
The CMS experiment uses a right-handed coordinate system, with the origin at the nominal interaction point, the $x$ axis pointing to the centre of the LHC, the $y$ axis pointing up (perpendicular to the LHC plane), and the $z$ axis along the anticlockwise-beam direction. The polar angle $\theta$ is measured from the positive $z$ axis and the azimuthal angle $\phi$ is measured in the $x$-$y$ plane.
The bore of the solenoid is instrumented with various particle detection systems. Charged-particle
trajectories are measured with a silicon pixel tracker with three barrel layers at radii between
4.4 and 10.2\unit{cm}, and a silicon strip tracker with 10 barrel detection layers that extend outward reaching
a radius of 1.1\unit{m}. Each system is completed by two endcaps, and provides angular coverage of $0 < \phi < 2\pi $ in
azimuth and $\abs{\eta} < 2.5$,  where the pseudorapidity $\eta$ is defined as $ \eta = -\ln [\tan(\theta/2)] $, with $\theta $  the polar angle of the trajectory of the particle with respect to the $z$ axis.
A lead-tungstate crystal electromagnetic calorimeter (ECAL) and a brass/scintillator hadron calorimeter (HCAL) surround the tracking volume and cover the region $\abs{\eta} < 3$.
The forward calorimeter  further extends the HCAL coverage  in the region $3 < \abs{\eta} < 5$,
improving the determination of the momentum imbalance.

Muons are measured and identified in gas-ionisation detectors embedded in the steel flux return yoke outside the solenoid in the range $\abs{\eta} < 2.4$. The barrel region is covered by drift-tube chambers  and the endcap region by cathode strip chambers, each complemented by resistive plate chambers.
The detector is nearly hermetic, allowing for energy balance measurements in the plane transverse to the beam.
A two-level trigger system selects the most interesting pp collision events for use in physics analysis.
The first level  of the CMS trigger system, composed of custom hardware processors, uses information from the calorimeters and muon detectors to select the most interesting events in a fixed time interval of less than about 3.2\mus. The second level, known as the  high-level trigger (HLT), is a processor farm that further decreases the event rate from around 100\unit{kHz} to around 300\unit{Hz}, before data storage.
A more detailed description of the CMS detector can be found elsewhere~\cite{CMS_detector}.

\section{Data and simulated samples}
\label{sec:scamples}

The measurements presented in this paper are performed using events from proton-proton collisions at a centre-of-mass energy of 7\TeV,
collected by the
CMS detector in 2011. The data correspond to an integrated luminosity of
$5.0 \pm 0.1\fbinv$~\cite{errorlumi}.
In order to account for effects of detector resolution and acceptance,
as well as to estimate the contribution from background processes that can
satisfy the
\ttbar event selection criteria, simulated event samples based on
Monte Carlo (MC) event generator programs are used.

A \ttbar signal sample was generated using the \MADGRAPH  generator version 5.1.1
\cite{MADGRAPH}  with matrix elements having up to three
extra partons in the final state
and an assumed top-quark mass of 172.5\GeVcc.
\MADGRAPH is interfaced with
\PYTHIA generator version 6.424~\cite{PYTHIA} to simulate hadronization and parton fragmentation and  with  \TAUOLA program version 27.121.5~\cite{tauola}
to simulate $\tau$-lepton decays.
The helicity fractions used as a SM reference are the LO predictions for $m_\cPqt=172.5$\GeVcc:
$F_0=0.6902, F_L=0.3089, F_R=0.0009$,  consistent with next-to-leading-order (NLO) and NNLO 
 expectations~\cite{WHelFracLO,Czarnecki:2010gb}.

Single-top quark events in $t$ and tW (or W-boson associated) channels  were
generated using \POWHEG program version 1.0~\cite{POWHEG} and \PYTHIA interfaced with  \TAUOLA.
Other relevant background processes, such as W boson and Drell--Yan production accompanied by multiple
jets, were simulated using \MADGRAPH.
In all LO simulations, the parton distribution function (PDF) set
CTEQ6L1~\cite{cteq} is used. The \POWHEG simulations use the NLO set
CT10~\cite{CT10}.

The effect of multiple proton-proton collisions occurring within the same bunch crossing (pileup events)
is taken into account in the simulation
and matches the pileup distribution observed in data.

\section{Event reconstruction and selection}
\label{sec:selection}

A set of requirements is applied to all samples, selecting candidate events
compatible with the topology of \ttbar  production. Almost all top quarks
decay into a W boson and a b quark. In the decay modes considered for
this study,
one of the W bosons decays hadronically into two jets and the other W boson decays
leptonically into an electron or muon
 and a neutrino. Hence, final states
containing a muon or electron
and at least four jets
are selected for further consideration.

Top-quark decay products are reconstructed using the particle-flow (PF) algorithm
described in detail in refs.~\cite{CMS-PAS-PFT-09-001} and~\cite{CMS-PAS-JME-10-003}.
The particle-flow event reconstruction identifies and measures the properties of
each particle, using an optimised combination of the information from
all subdetectors. The energy of photons is directly obtained from the ECAL measurement, corrected for zero-suppression effects. The energy of electrons is determined from a combination of the
momentum of the track originated at the main interaction vertex,
the corresponding ECAL cluster energy, and the energy sum of all bremsstrahlung photons attached to the track. The energy of muons is obtained from the corresponding track momentum. The energy of charged hadrons is determined from a combination of the track momentum and the corresponding ECAL and HCAL energy, corrected for zero-suppression effects, and calibrated for the nonlinear response of the calorimeters. Finally, the energy of neutral hadrons is obtained from the corresponding calibrated ECAL and HCAL energy.
The missing transverse energy \ETm
is defined as the magnitude of the transverse momentum imbalance, $\ptvec^\text{miss}$, which is the negative of the vectorial sum of the transverse momenta \pt
of all the particles reconstructed with the particle-flow algorithm.
Tracks belonging to the primary or secondary vertices of the most
energetic pp interaction are retained, while particles identified as coming from
pileup interactions  are removed from the event.

Events in the muon+jets channel were selected by a trigger that required at least one isolated,
high-momentum muon with a HLT \pt threshold varying
between 17 and 24\GeVc for the running period used in this analysis.
Events containing a
muon, well measured in the silicon
tracker and identified in the muon chambers with
$\pt>25$\GeVc and pseudorapidity $\abs{\eta} < 2.1$, are then selected offline. The track associated with the
muon candidate is required to have a minimum number of hits in the silicon tracker,
to be consistent with originating from the beam spot, and to have a high-quality global
fit including a minimum number of hits in the muon detector.
More details on the muon quality requirements are given in \cite{PAS-TOP-10-002}.

The trigger used to collect events in the electron+jets channel required at least
one isolated electron with  $\pt> 25\GeVc$, accompanied by at least three jets
with $\pt>$ 30\GeVc. Electron candidates
are reconstructed from clusters of energy deposits in the electromagnetic calorimeter,
which are then matched to hits in the silicon tracker~\cite{ref:electrons}.
Electrons are  identified by using shower shape and track-cluster matching variables.
Offline, events with exactly one electron candidate with $\pt>30$\GeVc and $\abs{\eta}<2.5$ are
selected. Events having electron candidates  in the transition region between the barrel and endcap calorimeters,
$1.44 <\abs{\eta}< 1.56$, are excluded because reconstruction in this region is degraded due to
additional material there. The electron track must lie within 0.02\unit{cm} of the primary vertex in the plane
transverse to the beam axis. Additionally, the background due to electrons
from photon conversions is reduced by rejecting tracks
with missing hits in the inner tracker layers or
that are near a track with an opposite charge and a similar polar angle.

Events with an additional muon with $\pt > 10$\GeVc or 
and additional electron with $\pt > 15$\GeVc are vetoed in order to
reject backgrounds from dileptonic
\ttbar and Drell--Yan events.

To reduce backgrounds further, muons and electrons are required to be
prompt and are therefore typically well isolated from the rest of the
event.  This is achieved via an offline particle-flow-based relative isolation
(PFIso) algorithm, which is defined as the sum of the transverse momenta
over all charged hadrons, neutral hadrons, and photons reconstructed inside
a cone of radius $\Delta R = \sqrt{(\Delta\eta)^2+(\Delta\phi)^2} = 0.3$,
centered around the lepton (muon or electron) and divided by the lepton
transverse momentum.  In top-quark decays, where the b jet coincidentally
overlaps with the prompt lepton, $\cos\theta^*$ takes values close to $-1$.
The offline lepton isolation requirement therefore has the effect of
reducing the signal selection efficiency near $\cos\theta^*\approx -1$.
In addition, both the muon and electron online triggers impose loose
isolation criteria.  Hence, selected muons (electrons) are required to have
$\text{PFIso} < 0.125\,(0.100)$.  These values are chosen to be tight enough
to provide a high trigger efficiency, yet loose enough to maintain
reasonable signal efficiency near $\cos\theta^*\approx -1$.  In order to
mitigate effects of pileup, a correction based on the average energy
density in the event is applied to the electron isolation.

Jets are reconstructed using the anti-\kt clustering algorithm~\cite{antikt_jet_algo,fastjet},
with  a distance parameter of 0.5, applied to the entire list of reconstructed
particles that are not identified as isolated muons or electrons in the event.
The resulting uncalibrated jet momenta are found to be within 5\% to 10\% of
the true momenta over the whole \pt spectrum and detector
acceptance.
Charged particles not associated with the primary vertex are explicitly removed from the jet, as stated above. A
residual correction is applied to
account for the energy of any extra neutral particles arising from pileup interactions.
  Further residual jet energy calibrations are derived from simulation and
corrected for any discrepancies with data using \textit{in situ} measurements
of object balancing in both dijet events (to render the jet response in
pseudorapidity uniform) as well as photon+jet events (to provide the absolute
jet energy scale)~\cite{CMSjes}.
Additional selection criteria are applied to remove spurious events with identified noise patterns in the HCAL.
Calibrated jets with $\pt > 10$\GeVc are used to correct the scale
of \ETm~\cite{CMSmet}.
Jet candidates from the top-quark decay
are required to have calibrated $\pt > 30$\GeVc and  $\abs{\eta} < 2.4$.
Events with less than four jets passing the
above mentioned top-quark decay product criteria are not used in the analysis.

To reduce the QCD multijet background, the transverse mass, \MT, of the leptonically decaying W boson,
is required to be greater than 30\GeVcc, where $\MT = \sqrt{\smash[b]{2 \pt \ETm(1-\cos(\Delta\phi))\times 1/c^{3} }}$
and $\Delta\phi$ is the angle in the $x$-$y$ plane between the direction of the
lepton and  $\ptvec^\text{miss}$.
In events where the top-quark pair decays dileptonically and one lepton escapes detection, the \MT variable can assume very high
values. Background events from this process are rejected by requiring $\MT < 200$\GeVcc.

Due to its relatively high rate and similar final state topology, the main remaining background source for this analysis is the production of several jets in association with a W boson that decays leptonically.  This background source can be reduced, and the QCD multijet background even further suppressed, by requiring that at least two of the selected jets be identified as b jets.  A high-efficiency tagging algorithm, known as the combined secondary-vertex (CSV) algorithm~\cite{ref:btag}, is used to separate jets originating from light quarks (or gluons) and heavy quarks, \ie charm or bottom quarks.  Jets are first divided into categories according to the probability of reconstructing a secondary vertex and its quality. Then, within each category, several variables including the three-dimensional signed impact parameter significance, secondary vertex mass, fractional charge, and charged particle multiplicity are used to form a likelihood that discriminates light-flavour jets from heavy-flavour jets. A selection working point is chosen so that the efficiency to identify a b jet is high (nearly 70\%), while the probability that a light-flavour jet is mistaken as a b jet is small (about 1\%). Requiring that there be two b-tagged jets in the event reduces the remaining QCD multijet background to negligible levels (less than 0.4\%).

Trigger, lepton identification, and lepton isolation efficiencies are estimated with a  \textit{tag-and-probe}
method~\cite{wzxs} using leptons from a sample of events containing
$\cPZ\to\ell^+\ell^-$ decays. The efficiencies
are computed for the $\cPZ$-boson events both in data and simulation, as a function of the lepton
\pt and $\eta$. The overall efficiencies in the typical \pt and $\eta$ ranges of the selected leptons are 80\% for muons and
70\% for electrons. The ratio
of the efficiencies in data and simulation, $\epsilon^{\mathrm{DATA}}_\ell/\epsilon^{\mathrm{MC}}_\ell$, is used as a scale factor to correct the
simulated samples.
Likewise, simulated samples are scaled to account for any differences
in b-tagging efficiencies between data and
simulation according to the ratio of efficiencies $\epsilon^\mathrm{DATA}_\text{b-tag}/\epsilon^\mathrm{MC}_\text{b-tag}$, which is determined as a function of the  \pt of the b jet~\cite{ref:btag}.

The number of data events reconstructed with these selection criteria
is 9268 in the muon channel and 6526 in the electron channel.
Comparisons between data observations and the SM expectations are presented in section~\ref{sec:datamc}.

\section{Top-quark reconstruction}
\label{sec:topreco}

Once events are selected according to the criteria described in section \ref{sec:selection}, the reconstruction of each top-quark candidate proceeds by testing all selected jets,  $\ptvec^\text{miss}$, and the lepton for their compatibility with decay products of the hadronic branch $(\cPqt\to {\cPqb\PW}\to { \cPqb \Pq\Paq^\prime})$ and the leptonic branch $(\cPqt\to {\cPqb\PW}\to \cPqb\ell\nu)$.
The initial value for the neutrino momentum is set to $\vec{p}^{\,\nu} = (\ptvec^\text{miss}, p^\nu_z )$, where $p^\nu_z$ is determined by requiring that the invariant mass of the neutrino and lepton be equal to the W-boson mass, which is assumed to be 80.4\GeVcc~\cite{pdg}.
For each possible neutrino solution and jet assignment to either the leptonic branch or hadronic branch, a $\chi^2$ is built, containing the following terms:
\begin{align}
\label{eq:chi2}
\chi^2_\text{comb} =& \left(\frac{m_\cPqt-m_\cPqt^\text{ref}}{\sigma_{m_\cPqt}}\right)^2 +
\left(\frac{m_{\cPaqt}-m_\cPqt^\text{ref}}{\sigma_{m_{\cPaqt}}}\right)^2 + \left(\frac{M_\PW^\text{lep}-80.4}{\sigma_{M_\PW^\text{lep}}}\right)^2 +  \left(\frac{M_\PW^\text{had}-80.4}{\sigma_{M_\PW^\text{had}}}\right)^2
\\
&-\sum_{i=1,4}2\ln p_i(\text{disc}|f), \nonumber
\end{align}
where $m_\cPqt$, $m_{\cPaqt}$, $M_\PW^\text{lep}$ and $M_\PW^\text{had}$ are the reconstructed
invariant masses for a given combinatorial assignment
of four jets to the final-state
particles in the $\ttbar$ decay.
The reference value for the top-quark mass $m_\cPqt^\text{ref}$ is taken to be 173.3\GeVcc~\cite{topmass}
for data and 172.5\GeVcc  for the simulated samples.
The term $p_i(\text{disc}|f)$ is  the probability for the $i$-th jet to have
flavour $f$ (either a b jet from the direct top-quark decay or a  jet from
the hadronically decaying W boson), given its measured value from the CSV tagger discriminant (see section \ref{sec:selection}).
Since the top-quark and W-boson reconstructed masses are dominated by experimental resolution effects,
the parameters $\sigma_{m_{\cPqt,\cPaqt}}$ and $\sigma_{M_\PW^\text{lep, had}}$ in eq.~(\ref{eq:chi2})
are approximated as Gaussian widths, which are determined from simulation.

In the process, a constrained kinematic fit is performed, which leads to an improved determination of the unmeasured neutrino momentum component $p^\nu_z$, or in some
cases a valid physical solution to be found when the analytical solution is imaginary due to detector mismeasurements, and to a more accurate reconstruction of the $\ttbar$ system. The momenta of the measured jets and lepton are allowed to vary within their resolutions and are required to comply with the same kinematic constraints used in eq.~(\ref{eq:chi2}).
The $\cPqt$ and $\cPaqt$ candidates that are chosen for further analysis correspond to the particular configuration of lepton, neutrino, and jets that minimise $\chi^2_\text{comb}$.  Any additional jets passing the event selection criteria that are not chosen as one of the four jets belonging to the $\ttbar$ system are discarded and no longer used in the analysis.
Events for which the kinematic fit fails to find a real solution complying with the constraints are discarded.

Following the full event selection in simulation studies, including the requirement of at least two identified b jets, this top-quark reconstruction algorithm associates the correct jet to the leptonic top-quark decay branch in 71\% of all cases.  If instead one loosens the requirement to at least one b jet in the event, the fraction of correctly assigned jets to the leptonic branch decreases to 63\%.

\section{Background  estimation}
\label{sec:bkg}

The main source
of backgrounds in the analysis comes from top-quark pairs that decay into either
fully leptonic or fully hadronic modes, or into semileptonic \ttbar decays involving taus,
and that pass the $\ell$+jets
selections.
Other backgrounds are, in order of decreasing importance: single top-quark events, events from processes involving W-boson production with jets (W+jets) and Drell--Yan production with jets (DY+jets).
The normalisation of the \ttbar processes is determined by the fitting procedure described in section \ref{sec:fit}, and the predictions for single top-quark processes
are determined from simulation (see section~\ref{sec:scamples}).

The cross section for inclusive W+jets and DY+jets production could be poorly predicted
in the specific phase space region where those events become background for top-quark
production, corresponding to high jet multiplicities and events containing
heavy-flavour jets.  For this reason, the
normalisation of the background coming from
W+jets and DY+jets is determined using an approach partially based on the data.
Muons and electrons are treated separately, since important requirements on background rejection, such as lepton isolation,
are different.

\subsection{Normalisation and shape of DY production with jets}

The normalisation and shape of distributions from DY+jets production, are studied
using a data control sample defined by applying the same selection criteria
described in section \ref{sec:selection}, except that events are required to
contain an additional lepton of the same flavour and opposite charge. In
this case, the \MT variable is computed with the event \ETm and the highest-\pt lepton. Using a reference
cross section of 3048\unit{pb}~\cite{fewz} for the DY production
 decaying into pairs of muons, electrons, and taus with invariant mass above 50\GeVcc,
the normalisation of simulated samples for
DY+jets is found to agree with the data  within the statistical
uncertainty of about 10\%. The following systematic uncertainties are then
considered in the estimation of the normalisation scale factor $N_\text{data}/N_\text{simulation}$, between the amount of data and simulated events:
\begin{itemize}
\item the
PDFs~\cite{pdfs,pdf4lhc} used to generate the samples: $\pm$3.6\%;
\item the uncertainties in the jet energy scale: $\pm$12.7\%;
\item the uncertainties in the jet energy resolution: $\pm$2.6\%;
\item the uncertainty in the integrated luminosity~\cite{errorlumi}:  $\pm$2.2\%;
\item the difference observed on $N_\text{data}/N_\text{simulation}$ for events inside and outside a window of 30$\pm$20\GeVcc around the
$\cPZ$-boson mass,  where the DY+jets background is probed: $\pm$20.0\%.
\end{itemize}
\noindent Including all systematic sources, this leads to a total estimated uncertainty of 30.0\% for the DY+jets normalisation.

The shape of the \costh ~distribution from DY+jets background events is
 verified using this same control sample. The top-quark reconstruction
algorithm described in section \ref{sec:topreco} is applied, except that the
highest-\pt lepton is used in the kinematic fit.  The shapes
observed in data and simulation are found to be in agreement.

\subsection{Normalisation and shape of W-boson production with jets}

Due to the charge of the valence quarks in the colliding protons, more
positively ($\ell^+$) than negatively ($\ell^-$) charged leptons are
produced in pp collisions for single top-quark production and W+jets
events.  On the other hand, the amount of $\ell^+$  and $\ell^-$ produced in
DY+jets and \ttbar processes is, to a very good approximation, the same in
pp collisions.  Hence, by keeping the predicted cross section of single-top-quark
production fixed, the background contribution from W+jets production can be
determined from the charge asymmetry $N_+-N_-$, where $N_+$ ($N_-$) is the number of $\ell^+$ ($\ell^-$).  The total
number of W+jets events in the data sample is thus predicted to be
\begin{equation}
(N_+ + N_-)^{\PW+\text{jets}}_\text{predicted} = R_{\pm \mathrm{(MC)}}^\PW\times (N_+ -
N_-)_\text{data},
\label{eq:datadriven}
\end{equation}
where $R_{\pm \mathrm{(MC)}}^\PW = (N_+ + N_-)^{\PW+\text{jets}}_\mathrm{MC}/(N_+ -
N_-)^{\PW+\text{jets}}_\mathrm{MC}$ is estimated using simulated events and $(N_+ -
N_-)_\text{data}$ is the measured asymmetry in data.
The fixed contribution of single-top quark from simulation is
subtracted from the total charge asymmetry in eq. (\ref{eq:datadriven}) so
that $(N_+ - N_-)_\text{data} = (N_+ - N_-)_\text{data}^\text{total} - (N_+ -
N_-)_\mathrm{MC}^\text{single-top} $.

The predicted normalisation is found to be consistent
with the expectations from the simulation within relatively large statistical
uncertainties.
A total systematic uncertainty of 100\% is assigned to the normalisation of
the predicted W+jets background.  To test the effect of possible biases due
to the assumed background shape from simulation, the analysis is repeated
by dividing the data sample in bins of \costh. To increase the statistical
power of the test, the jet selection criteria are partially relaxed.
Within the precision of the test, the shape of the \costh~distribution
predicted from the simulation is found to be in agreement with the data,
and any possible bias is negligible compared with the normalisation
uncertainty considered.

\section{Determination of helicity fractions}
\label{sec:fit}

A fitting procedure based on a MC simulation reweighting technique
is used to simultaneously account for experimental resolution effects and
for the dependencies on the W-boson helicity fractions. Any new helicity
configuration can be obtained from the original configuration used in the
simulation via an algorithm that (re)weights each event according to the
generated, matrix-element level $\cos\theta^*$ values for each $\PW\to \ell\nu$
and $\PW\to \cPq\cPaq$ decay in the \ttbar simulated sample.

Because of the QCD production mechanism for \ttbar events, top quarks can
be considered as unpolarised on average, to a high degree of precision.
 Spin correlations between the two top quarks in the event do not modify
this picture.  This is because the average spin properties of the top quark
associated with the leptonic W-boson decay branch do not change after averaging
over the phase space variables of the other top quark (from the hadronic W-boson
decay branch) in the event. This scenario, which has been assumed in
past and recent W-boson helicity studies~\cite{AtlasPaper,CDFwhel,D0whel}, implies
that the phase space density for the reconstructed $\cos\theta^*$ variable
of each decay branch, $\rho(\cos\theta^*_\text{rec})$, can be decoupled from the
rest of the event. Therefore, at matrix-element (generator) level, we
assume the following dependence:
\begin{equation}
  \rho(\cos\theta^*_\text{gen}) \equiv
  \frac{1}{N}
  \frac{\rd{}N}{\rd\cos\theta^{*}_\text{gen}}=
  \frac{3}{8} F_L (1-\cos\theta^{*}_\text{gen})^2 +
  \frac{3}{4} F_0 \sin^{2}\theta^{*}_\text{gen} +
  \frac{3}{8} F_R (1+\cos\theta^{*}_\text{gen})^{2}, \label{eq::Reweighting::Theory}
\end{equation}
\noindent
where $(F_L,F_0,F_R)\equiv \vec{F}$ are the helicity fractions to be
measured, and $\theta^{*}_\text{gen}$ is the matrix-element level quantity for the helicity angle.
For normalisation reasons, the helicity fractions are
constrained to satisfy $F_L + F_0 + F_R = 1$. A new $\cos\theta^*_\text{rec}$
distribution for a particular configuration of helicity fractions $F_L,
F_0, F_R$ is then obtained by reweighting each fully simulated event by the
weight
\begin{equation}
W(\cos\theta^{*}_\text{gen};\vec{F}) = W_\text{lep}(\cos\theta^{*}_\text{gen};\vec{F}) \times W_\text{had}(\cos\theta^{*}_\text{gen};\vec{F}),
\end{equation}
where $W_\text{lep, had}$ is defined for the leptonic and hadronic branches respectively as:
\begin{align}
  W_\text{lep, had}(\cos\theta^{*}_\text{gen};\vec{F}) &\equiv
  \frac{\rho(\cos\theta^*_\text{gen})}{\rho^\mathrm{SM}(\cos\theta^*_\text{gen})} =  \nonumber \\
  &= \frac
  {\displaystyle
  \frac{3}{8} F_L (1-\cos\theta^{*}_\text{gen})^2 +
  \frac{3}{4} F_0 \sin^{2}\theta^{*}_\text{gen} +
  \frac{3}{8} F_R (1+\cos\theta^{*}_\text{gen})^{2}
  }
  {\displaystyle
  \frac{3}{8} F^\mathrm{SM}_L (1-\cos\theta^{*}_\text{gen})^2 +
  \frac{3}{4} F^\mathrm{SM}_0 \sin^{2}\theta^{*}_\text{gen} +
  \frac{3}{8} F^\mathrm{SM}_R (1+\cos\theta^{*}_\text{gen})^{2}
  }.
  \label{eq:Reweighiting:weightDef}
\end{align}
\noindent
In the expression above, $F^\mathrm{SM}_L, F^\mathrm{SM}_0, F^\mathrm{SM}_R$ are the
helicity fractions that correspond to the expected SM values and which
are used to generate the reference  simulated sample.
The new reweighted distribution automatically takes into account, by construction,
all detector resolution and acceptance effects, as described by the simulation.

The helicity fractions are measured by maximising a binned Poisson
likelihood function,
\begin{equation}
\mathcal{L}(\vec{F}) = \prod_\text{bin $i$} \frac{N_\mathrm{MC}(i;\vec{F})^{\displaystyle N_\text{data}(i)}}{(N_\text{data}(i))!}~\exp{(-N_\mathrm{MC}(i;\vec{F}))},
\end{equation}
\noindent
which is constructed using the numbers of observed ${N_\text{data}}(i)$
and expected ${N_\mathrm{MC}}(i;\vec{F})$ events in each $\cos\theta^*_\text{rec}$ bin $i$. The
numbers of expected events are given by:
\begin{align}
{N_\mathrm{MC}}(i,\vec{F}) & =  {N_\mathrm{BKG}}(i) + {N_{\ttbar}}(i;\vec{F}), \\
{N_{\ttbar}}(i;\vec{F}) & =  \mathcal{F}_{\ttbar}\left[\sum_{\text{\ttbar events}} W(\cos\theta^{*}_\text{gen} (i);\vec{F})\right]
\label{eq::Reweighting::WeightDef1}, \\
{N_\mathrm{BKG}}(i) & =  {N_{\PW+\text{jets}}}(i)+{N_{\mathrm{DY}+\text{jets}}}(i)+{N_\text{single-top}}(i)+\mathcal{F}_{\ttbar}\times N_{\ttbar\ \text{non-}\ell+\text{jets}}(i).
\end{align}

The normalisation parameter for the \ttbar component $\mathcal{F}_{\ttbar}$ is
not sensitive to the helicity fractions, but it does absorb a large
fraction of the experimental and theoretical systematic uncertainties in
the predicted rates. Uncertainties on the normalisation of backgrounds are
considered as a separate source of systematics.

Two types of fits are performed. In the fits denoted by 3D, the fractions
$F_0, F_L$, and the normalisation factor $\mathcal{F}_{\ttbar}$ (see eq.~(\ref{eq::Reweighting::WeightDef1})) are treated as free parameters, and
$F_R$ is determined by the constraint $F_R=1-F_0-F_L$.  In the fits denoted
by 2D, $F_0$ and $\mathcal{F}_{\ttbar}$ are taken as free parameters, setting
$F_R=0$, and solving for $F_L$ from the constraint $F_L=1-F_0$.  Within the
expected experimental uncertainties, the $F_R = 0$ condition is satisfied
both in the SM and in anomalous coupling scenarios where only $g_R$ is
different from zero.

We perform the measurements by fitting either the \costh~distribution from
the leptonic branch or the $\abs{\costhhad}$ distribution from the hadronic
branch. The two variables are fitted separately and not simultaneously, so
as to avoid any possible biases due to top-quark spin correlations. At
matrix-element level, $\abs{\costhhad}$ is only sensitive to the $F_0$ fraction
(or, alternatively, to the combination $F_L+F_R$), and is therefore only
used in the context of the 2D fits.

\section{Comparison between data and SM predictions}
\label{sec:datamc}

The agreement between data and expectation from the SM is extensively
investigated.  First, the normalisations of simulated samples involving
W+jets and DY+jets are determined and applied using
control data samples discussed in section~\ref{sec:bkg}.
Next, reference  cross sections 64.6\unit{pb}~\cite{ttch} for single-top-quark ($t$-channel), and 15.74\unit{pb}~\cite{ttw} for
W-boson-associated
single-top-quark processes, which correspond to approximated NNLO predictions,
and NLO \ttbar cross sections are used to scale  the respective simulated samples to an integrated luminosity of 5.0\fbinv.
Finally, the simulated samples involving top quarks
are corrected for both the presence of pileup and the efficiency
correction factors $\epsilon_{\ell}^\text{DATA}/\epsilon_{\ell}^\mathrm{MC}$ and
$\epsilon_\text{\cPqb-tag}^\text{DATA}/\epsilon_\text{\cPqb-tag}^\mathrm{MC}$.

Table \ref{tab:nevt} presents the number of data events observed in
the muon+jets and electron+jets channels which is compared
to the predictions
for the
signal (\ttbar) and background processes.  In the table, columns two and
four display the number of remaining events after applying the selection
criteria described in section \ref{sec:selection}.  Columns three and five
display the subsample  of these events where a \ttbar pair candidate is
found, reconstructed as described in the previous section.
The yields observed in data are in agreement with the
predicted yields expected from the SM.

\begin{table}[htp]
\begin{center}
\topcaption{
Number of events expected from  signal and background processes, together
with the observed number of events in data for both the muon+jets and
electron+jets channels.  The columns labelled as ``Selected" represent the
number of events passing the analysis selection criteria; the columns
labelled as ``KF" represent the fraction of those events containing a
reconstructed top-quark pair via a kinematic fit that has converged.
\label{tab:nevt}}
\begin{tabular}{|l|c|c||c|c|}\hline
                  & \multicolumn{2}{c||}{muon+jets}   &                    \multicolumn{2}{c|}{electron+jets}                \\ \cline{2-5}
 Process          	  &    Selected   &   KF    &   Selected    &   KF      \\ \hline\hline
Single top         	   &  372   &   319  &   265   &   223   \\
W+jets              	   &  351   &   282  &   234   &   203   \\
DY+jets             	   &  43    &   38   &   43    &   34    \\
\ttbar non $\ell$+jets 	   &  928   &   828  &   629   &   547   \\ \hline
Total bkg.         	   &  1694  &   1467 &   1171  &   1007  \\
\ttbar signal     	   &  7597  &   7321 &   5391  &   5179  \\ \hline
Total expected      	   &  9291  &   8788 &   6562  &   6186  \\ \hline
Data                	   &  9268    &   8772   &   6526    &   6135    \\ \hline
\end{tabular}
\end{center}
\end{table}

\begin{figure}[tp]
\centerline{ \includegraphics[width = 0.45\textwidth]{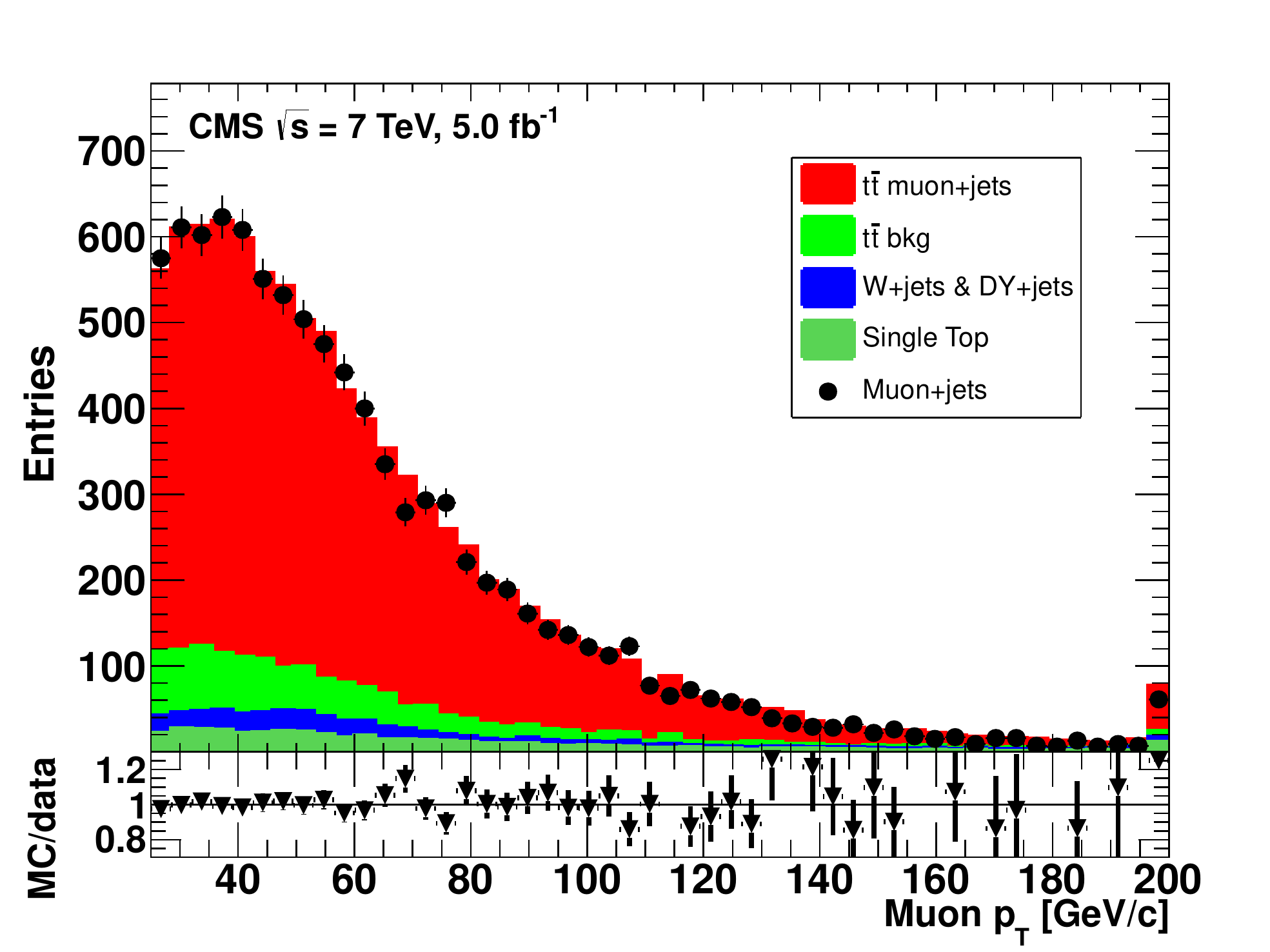}  \includegraphics[width = 0.45\textwidth]{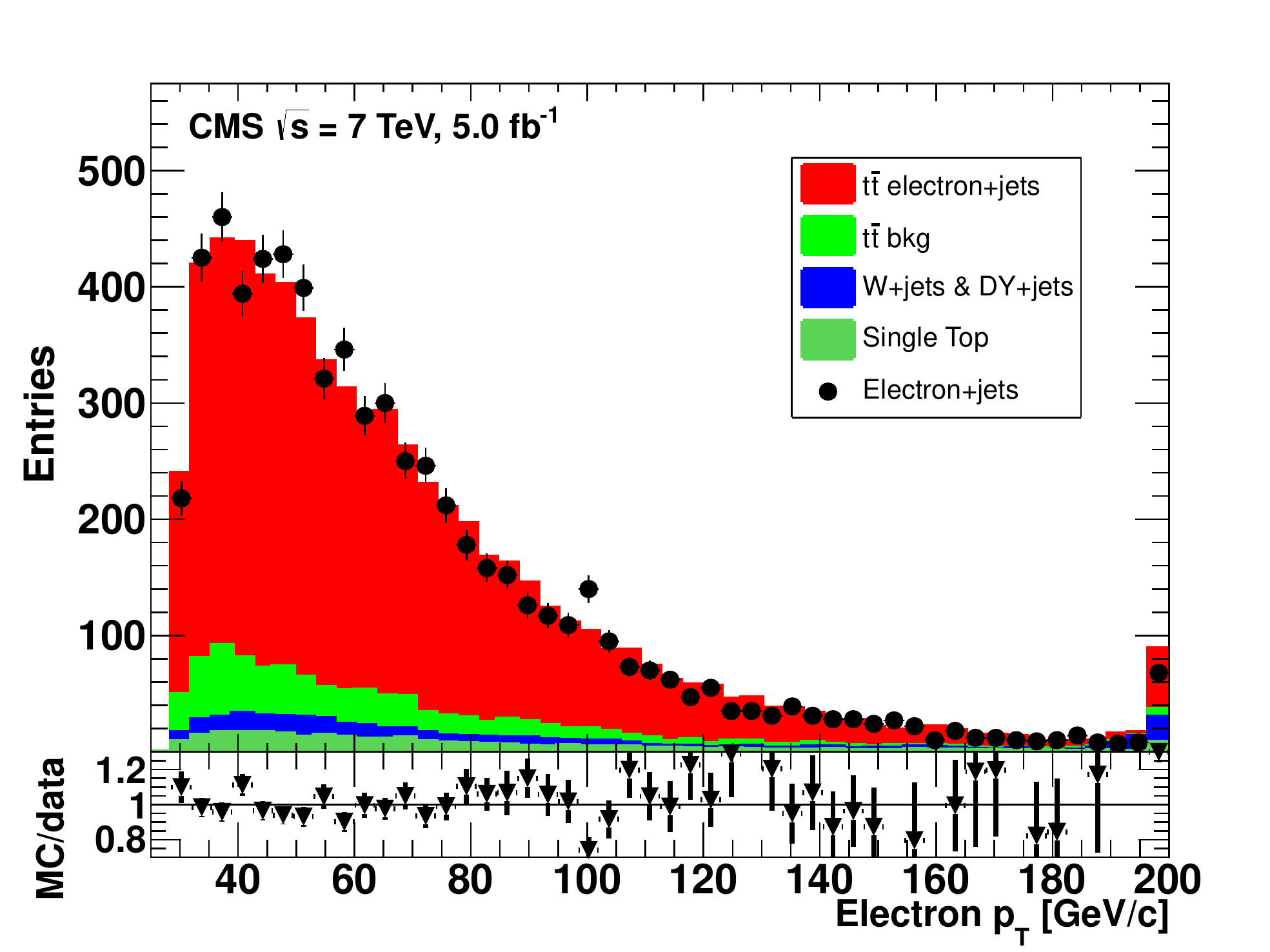} }

\centerline{ \includegraphics[width = 0.45\textwidth]{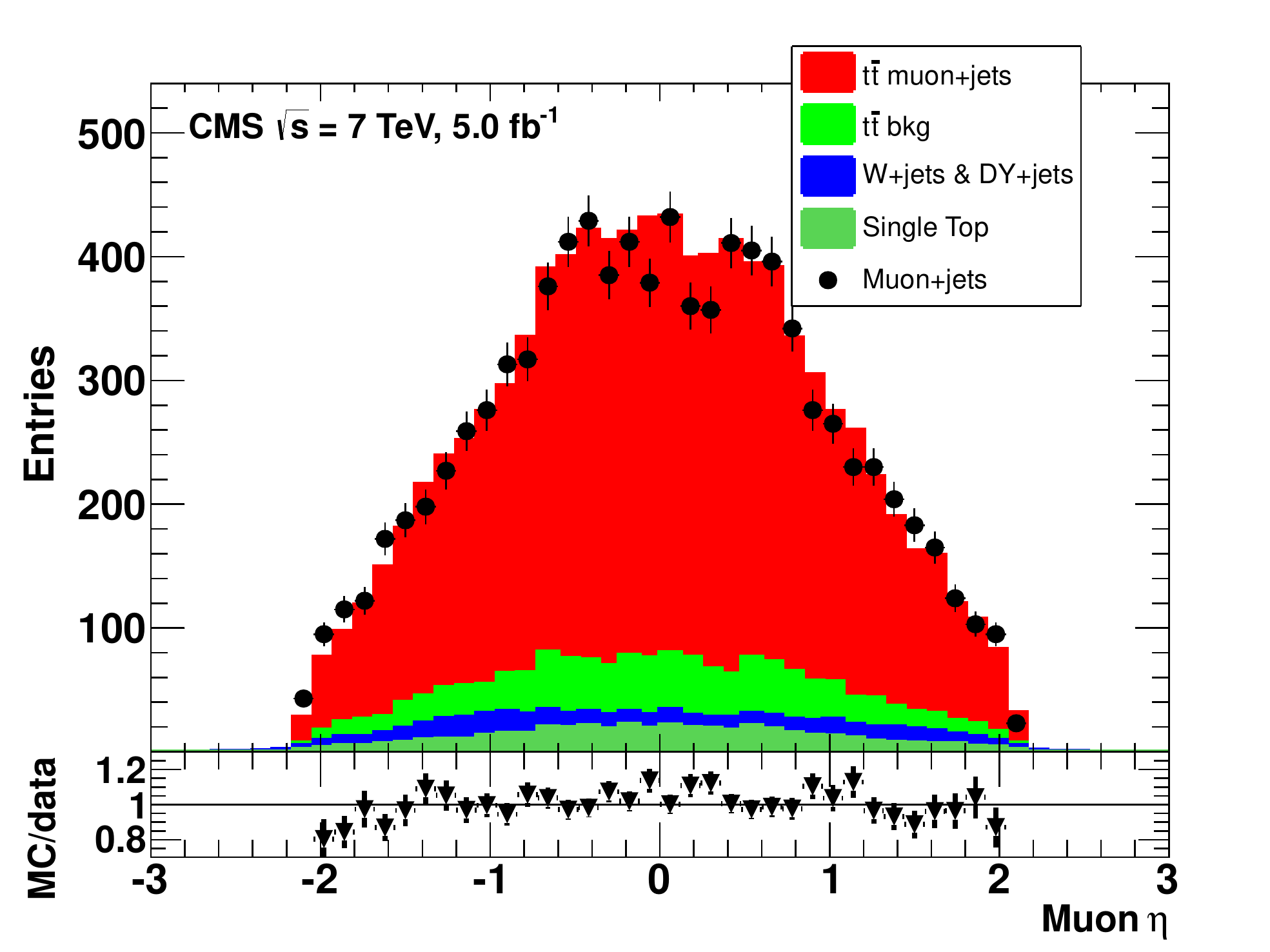}  \includegraphics[width = 0.45\textwidth]{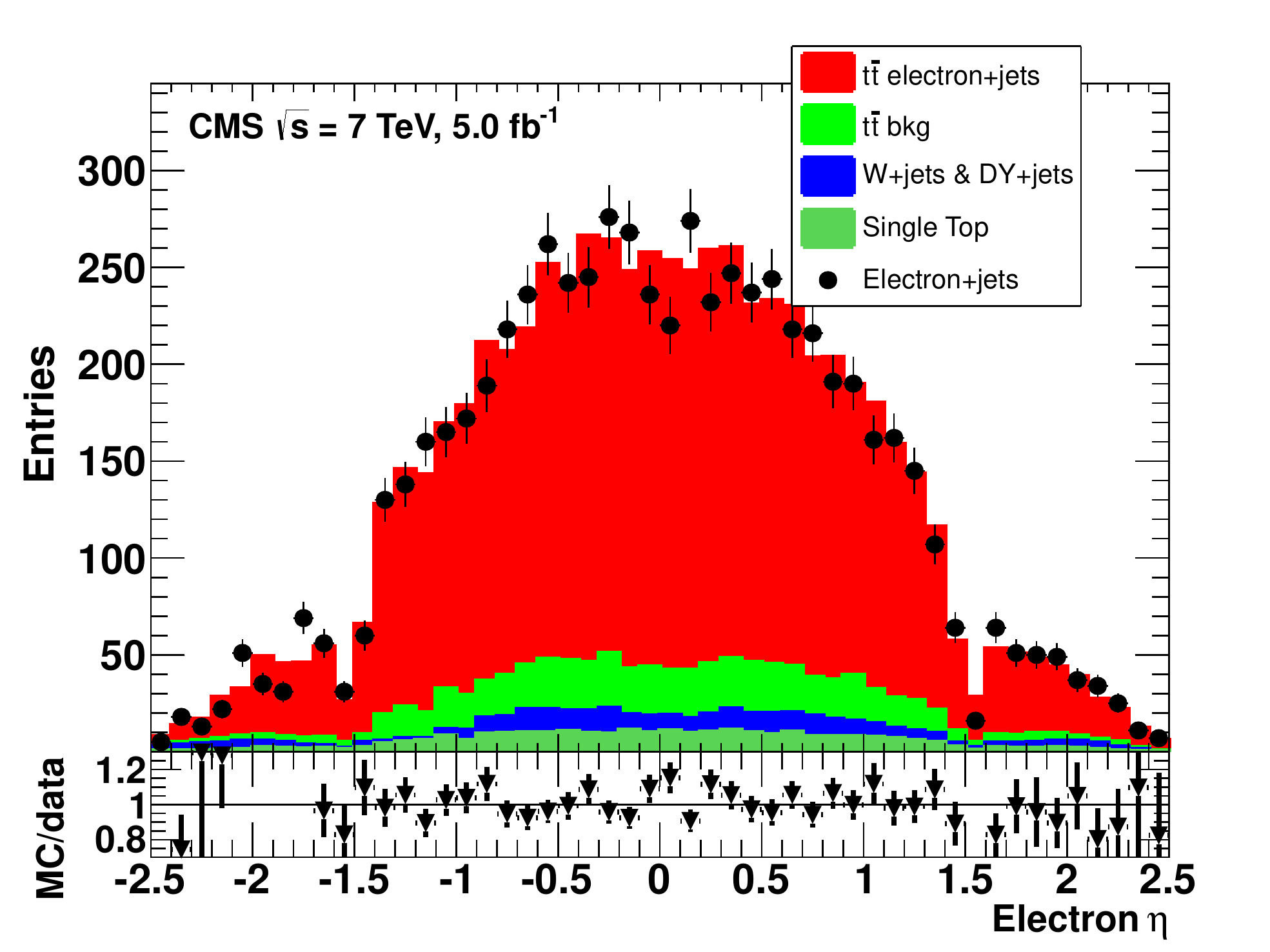} }

\centerline{ \includegraphics[width = 0.45\textwidth]{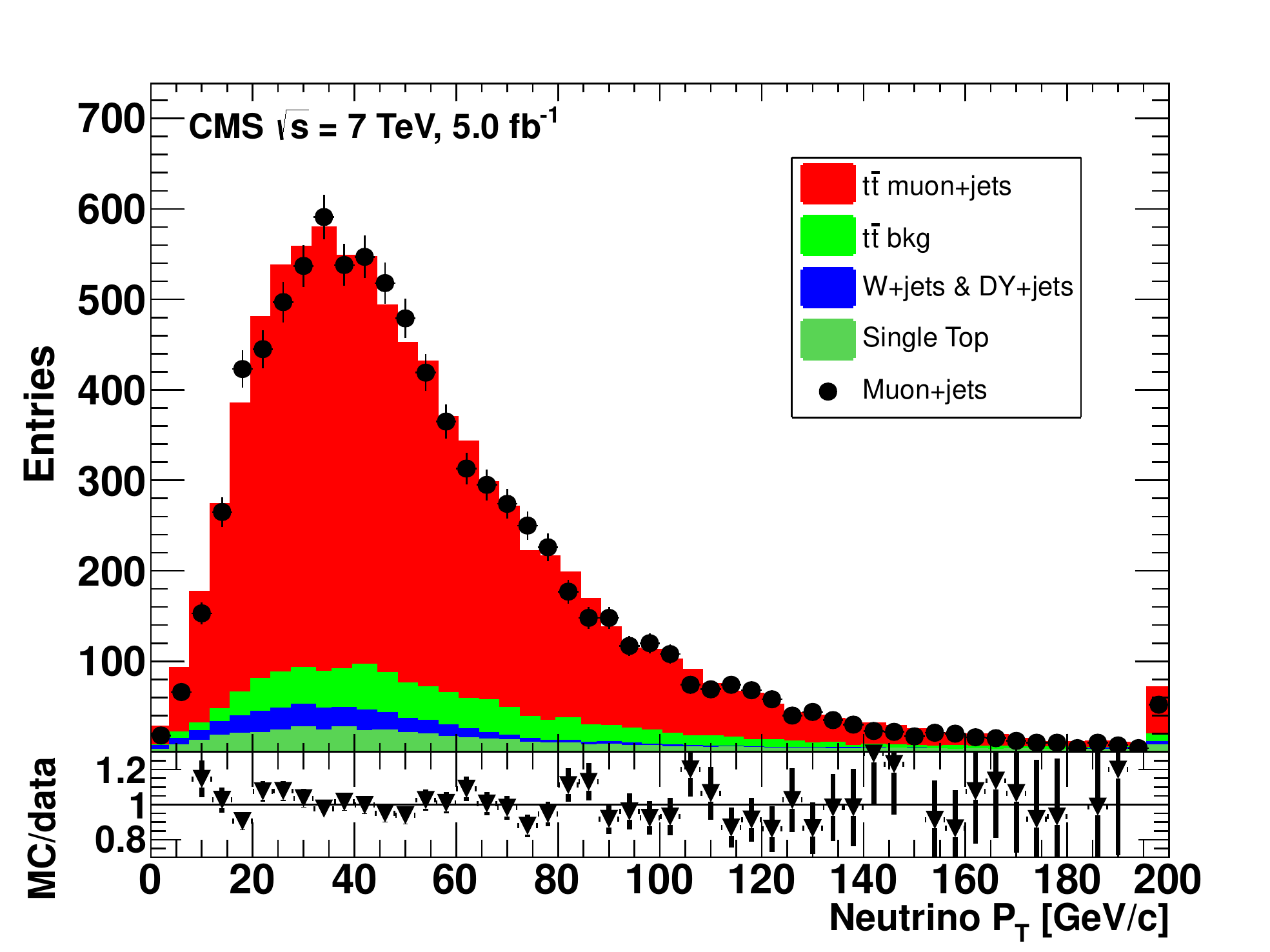}  \includegraphics[width = 0.45\textwidth]{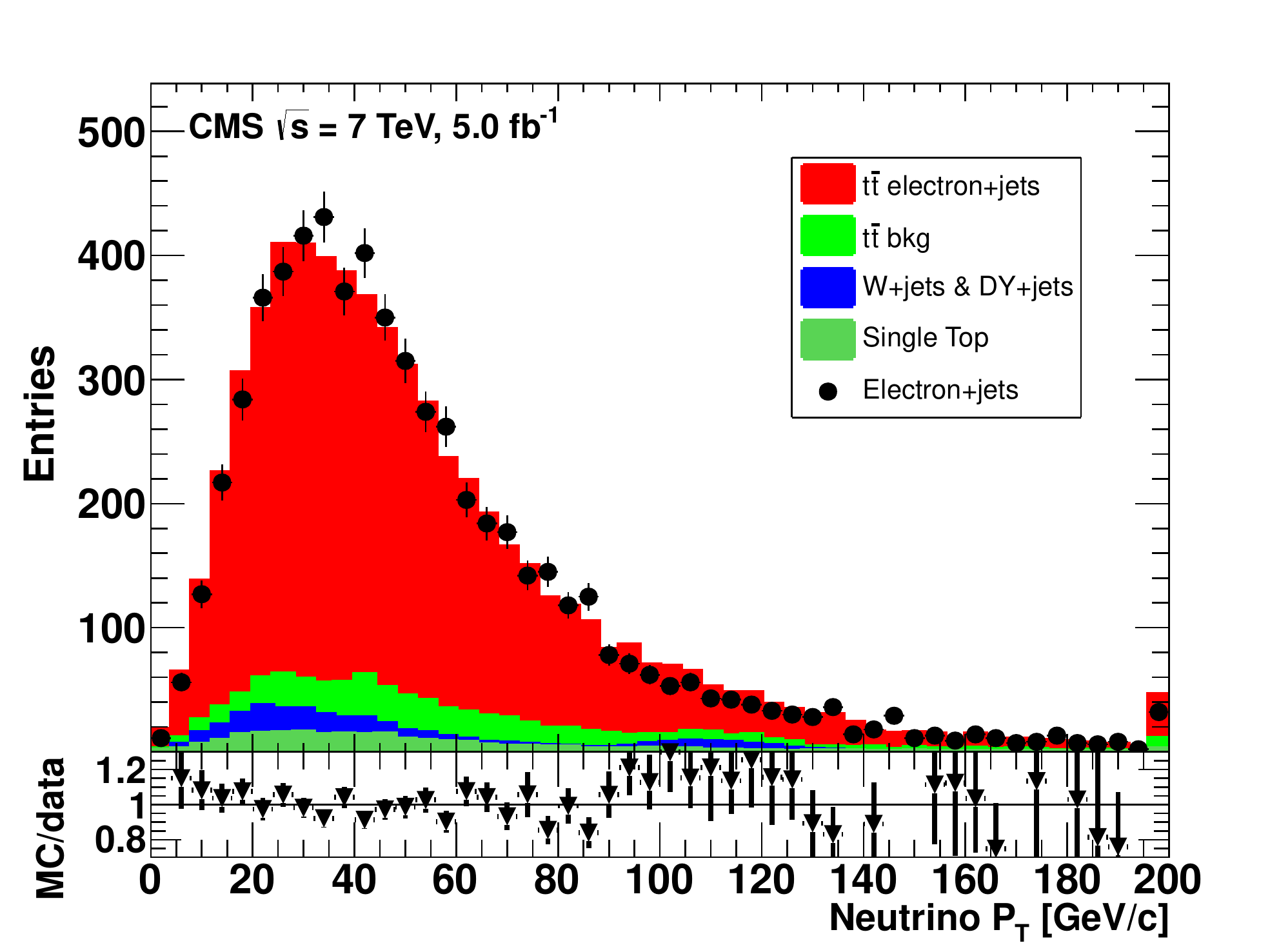} }
\caption{
Distributions of
data compared to SM predictions for
signal and expected backgrounds: charged lepton \pt (top), $\eta$ (centre) and
neutrino \pt (bottom)
for the muon+jets (left) and electron+jets (right) channels.  Data are displayed
as solid points, simulated \ttbar signal distributions as red histograms, and the
contribution from other background processes as coloured histograms.
Overflows are displayed in the last bin of each histogram.  At the bottom,
the ratio between prediction and data is displayed. Only statistical
uncertainties are shown.
 \label{fig:controlmu}}

\end{figure}

\begin{figure}[tp]
\centerline{ \includegraphics[width = 0.45\textwidth]{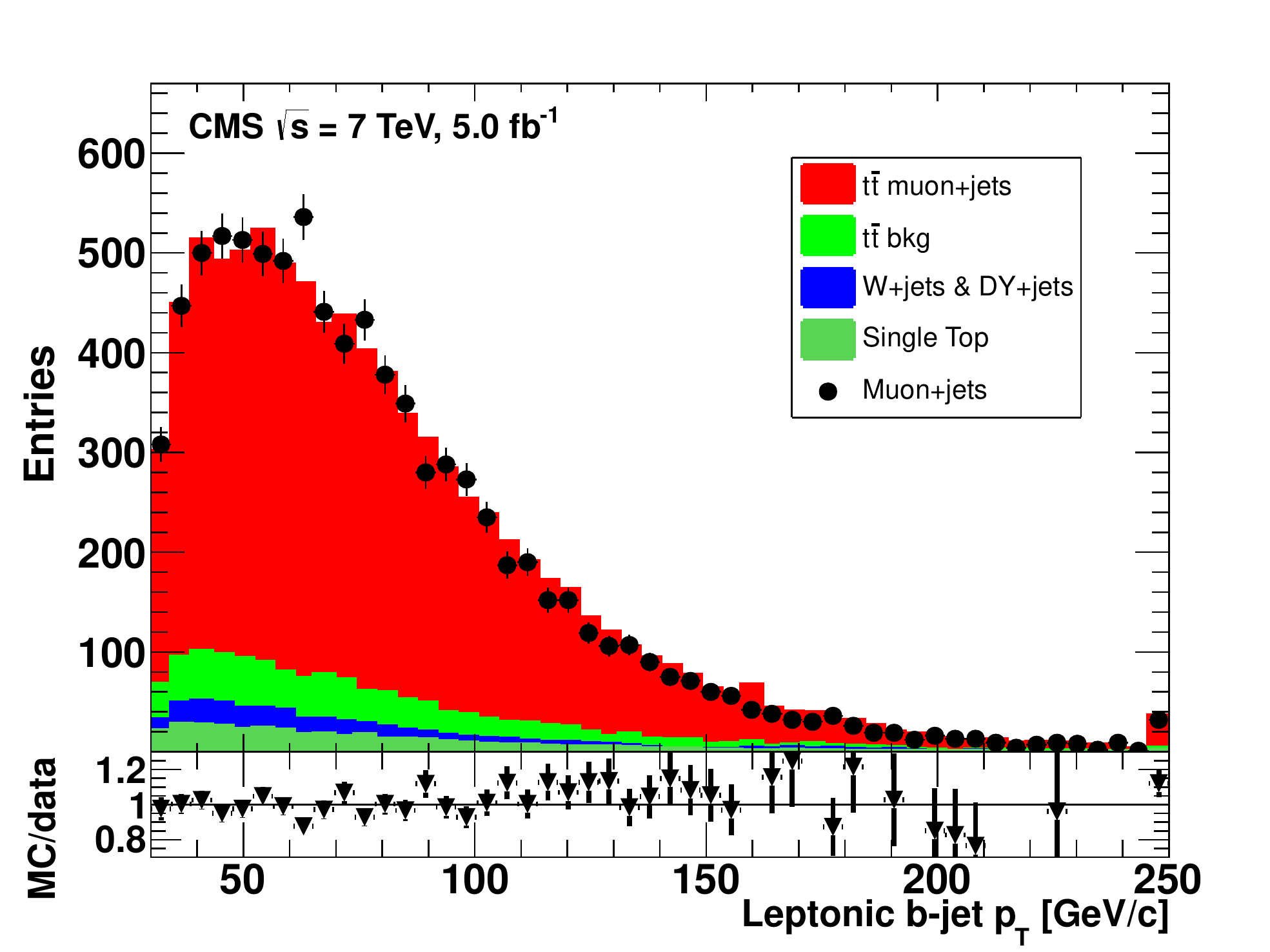}  \includegraphics[width = 0.45\textwidth]{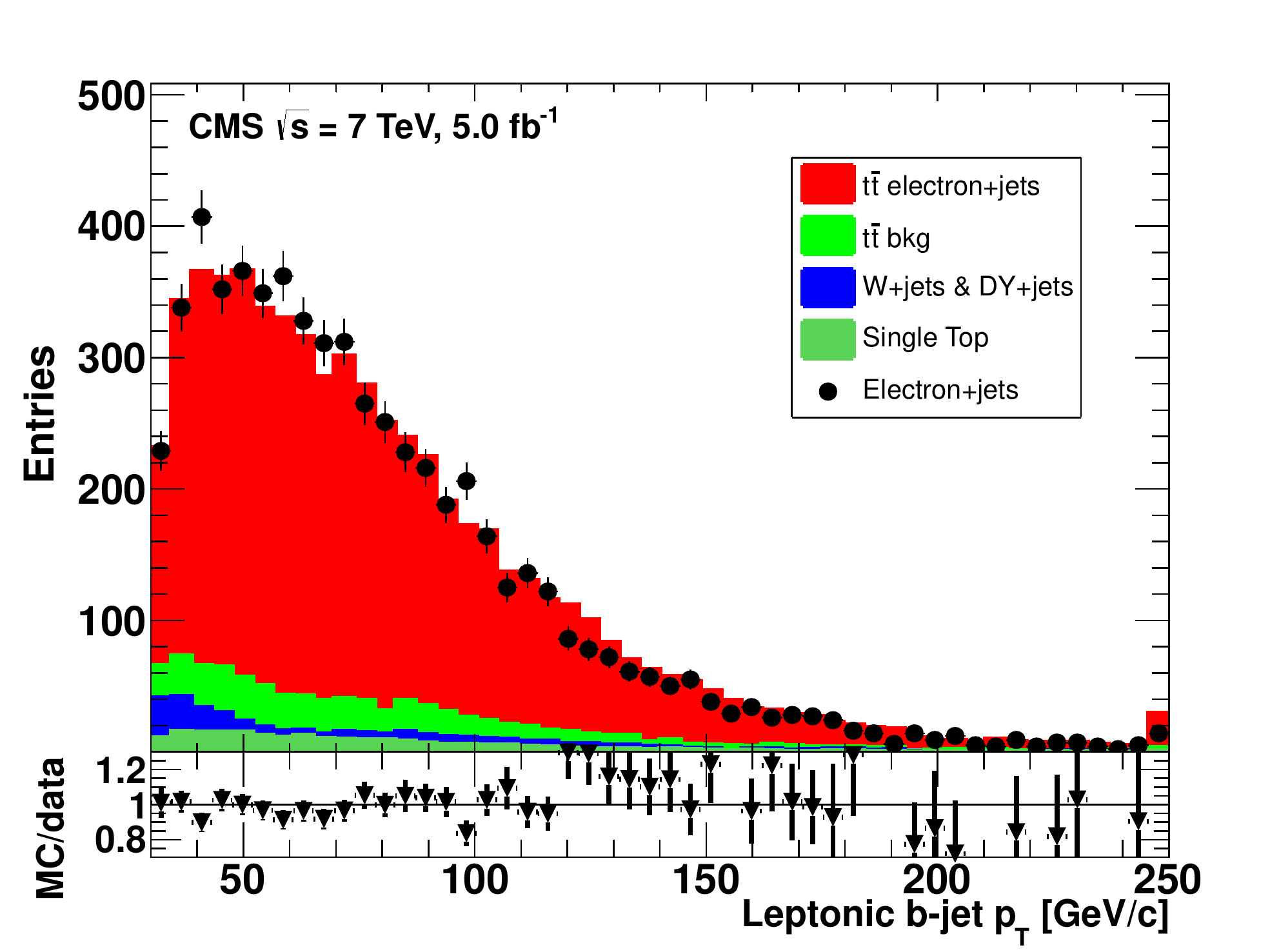} }

\centerline{ \includegraphics[width = 0.45\textwidth]{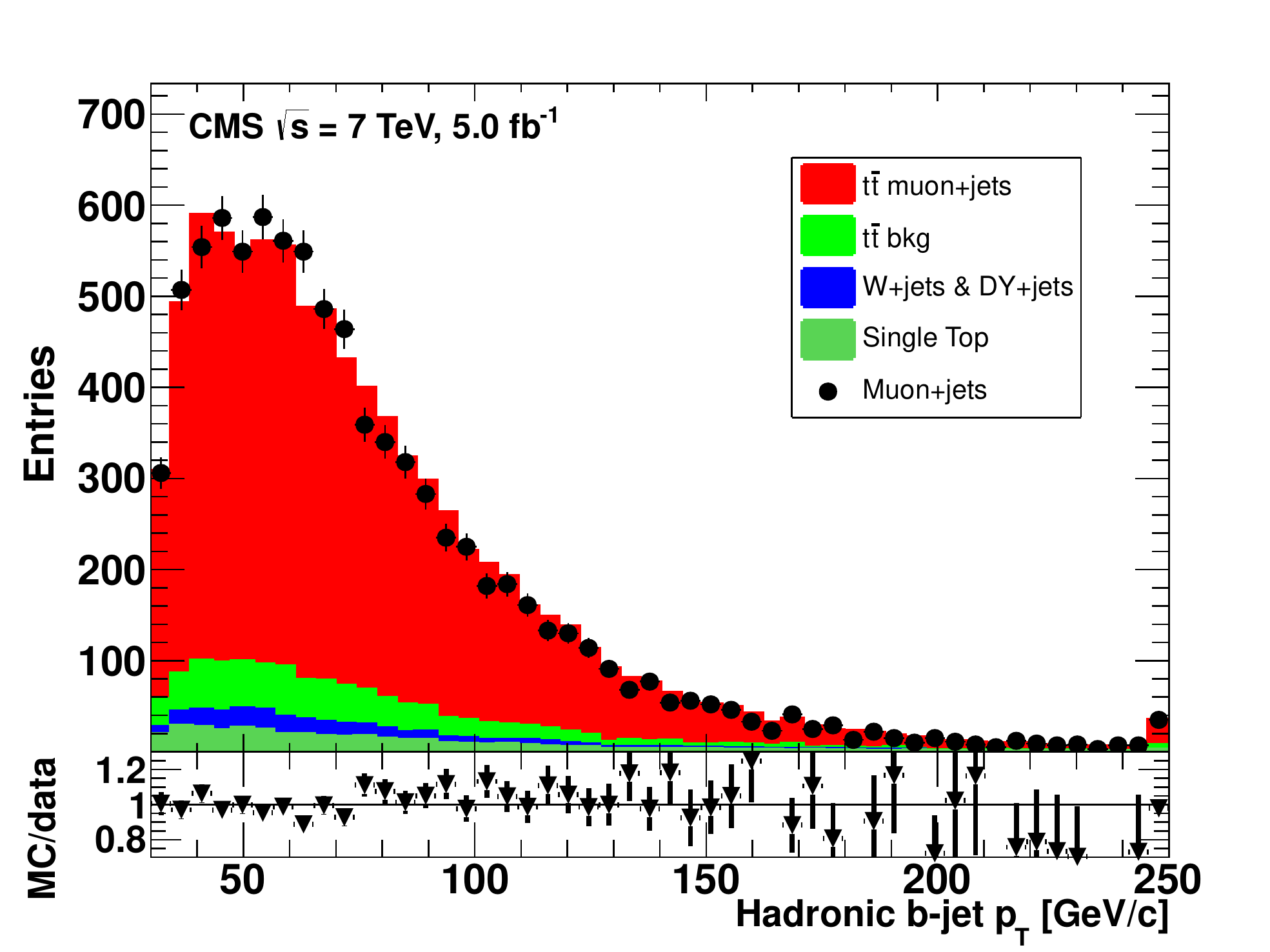}  \includegraphics[width = 0.45\textwidth]{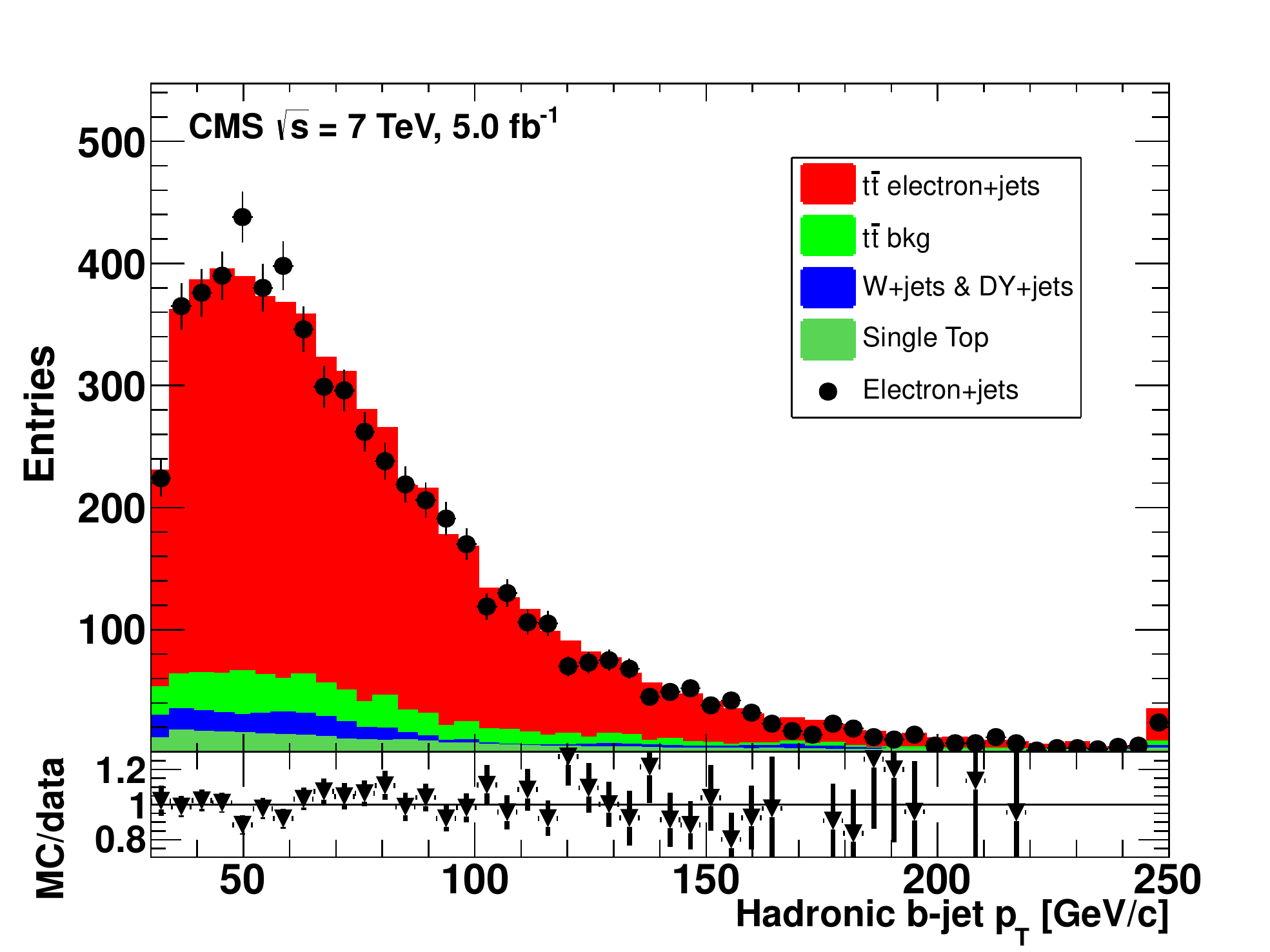} }

\centerline{ \includegraphics[width = 0.45\textwidth]{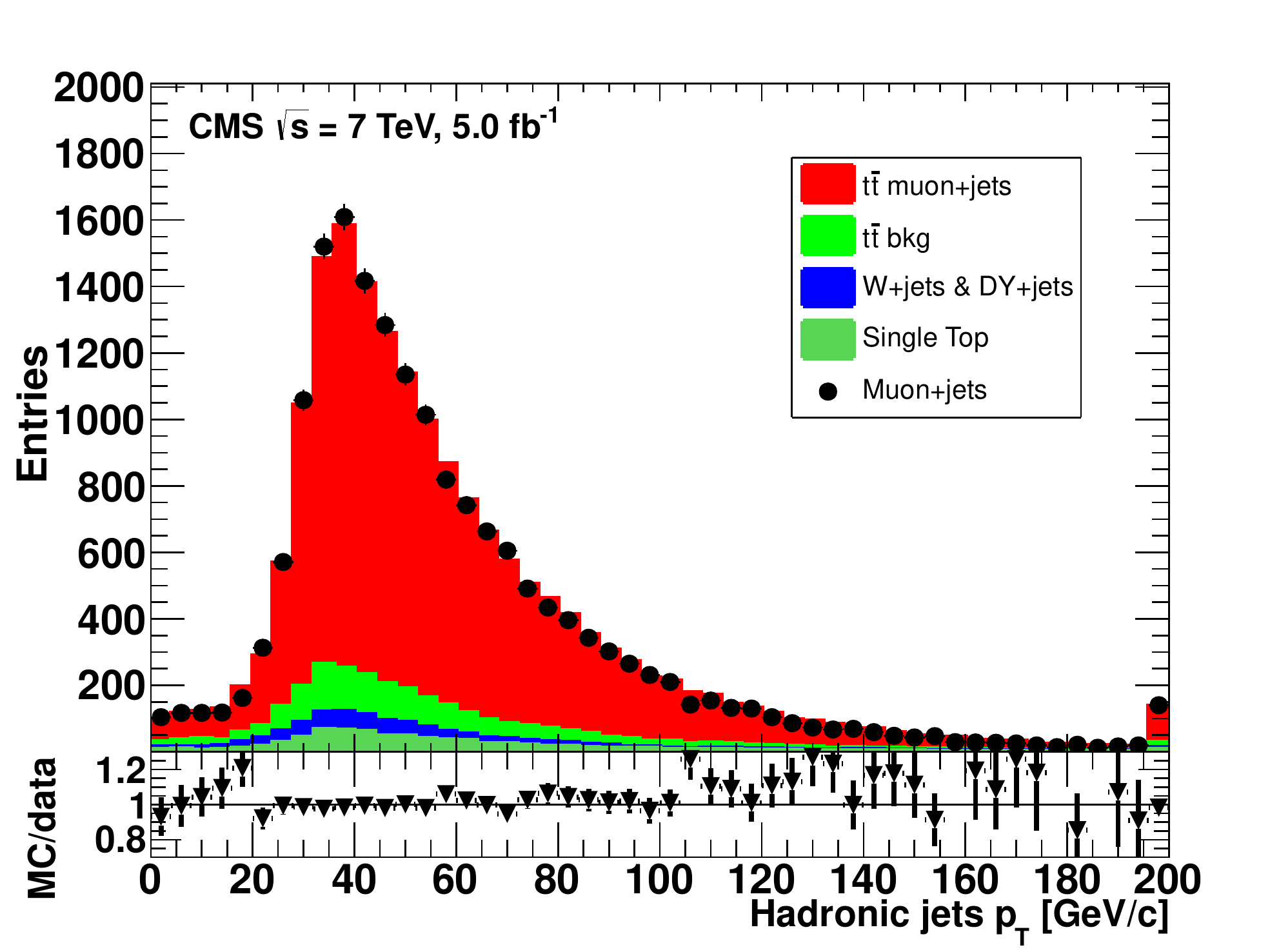}  \includegraphics[width = 0.45\textwidth]{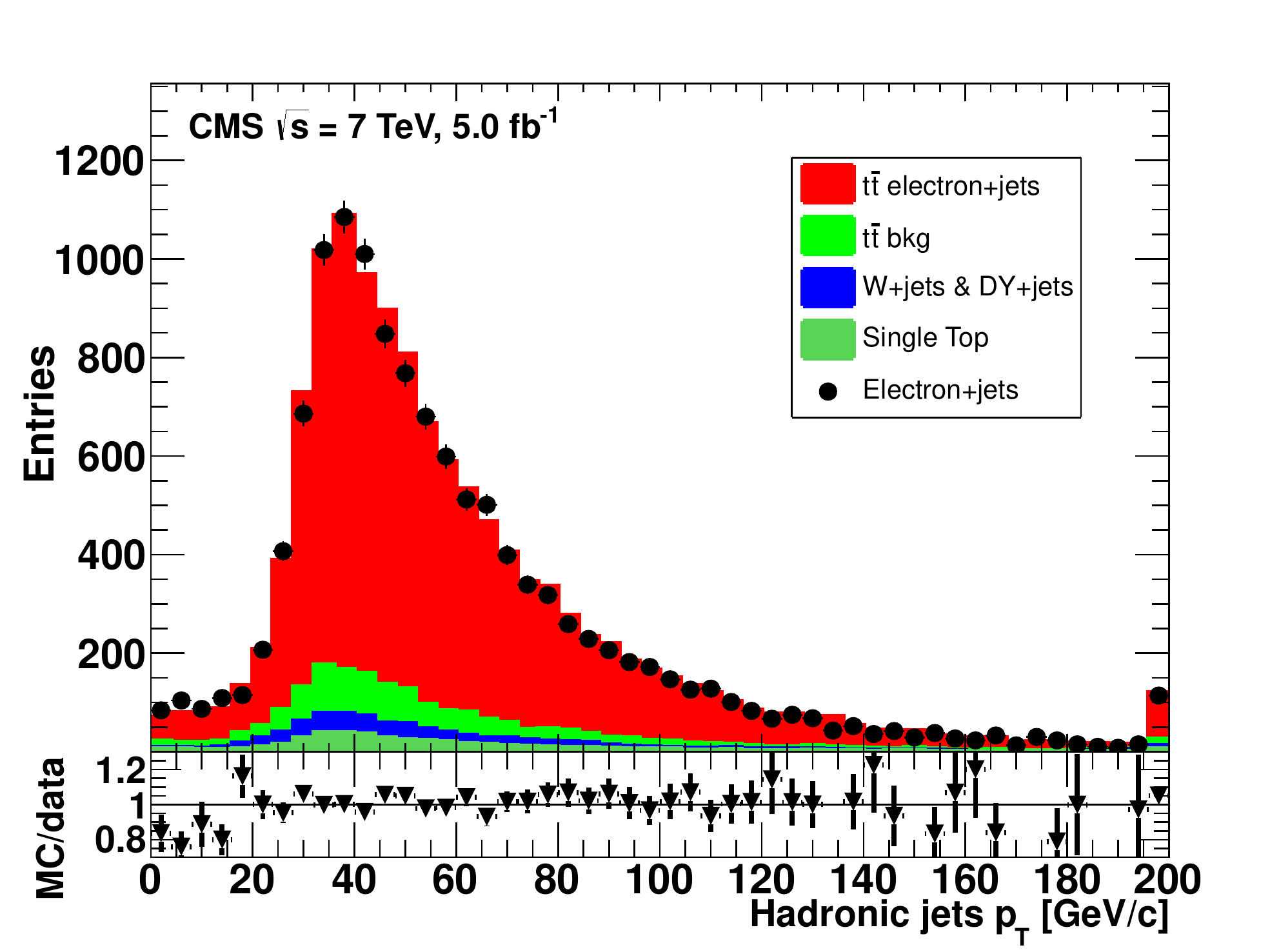} }

\caption{
Distributions of transverse momenta in data compared to SM predictions for
signal and expected backgrounds:  the
b jets identified in
the leptonic (top row) and hadronic (second row) branches, and the
jets from the hadronic W decay with two entries per event
(bottom row)
for the muon+jets (left) and
electron+jets (right) channels.  Data are displayed as solid points,
simulated \ttbar signal distributions as red histograms, and the contribution from
other background processes as coloured histograms. Overflows are
displayed in the last bin of each histogram.  At the bottom, the ratio
between prediction and data is displayed. Only statistical uncertainties
are shown.
\label{fig:jetNu}}
\end{figure}

\begin{figure}[t]
\centerline{ \includegraphics[width = 0.45\textwidth]{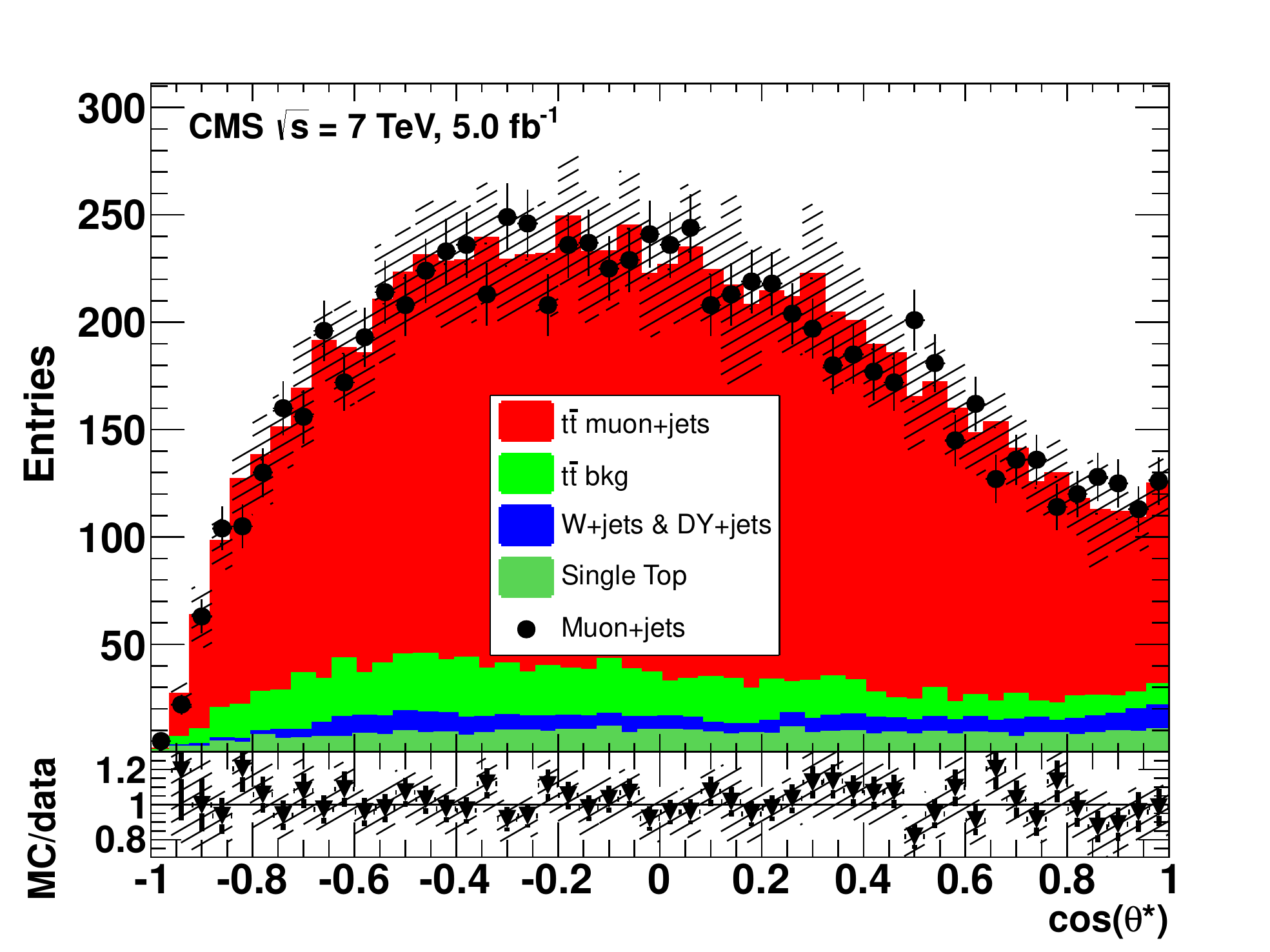}
\includegraphics[width = 0.45\textwidth]{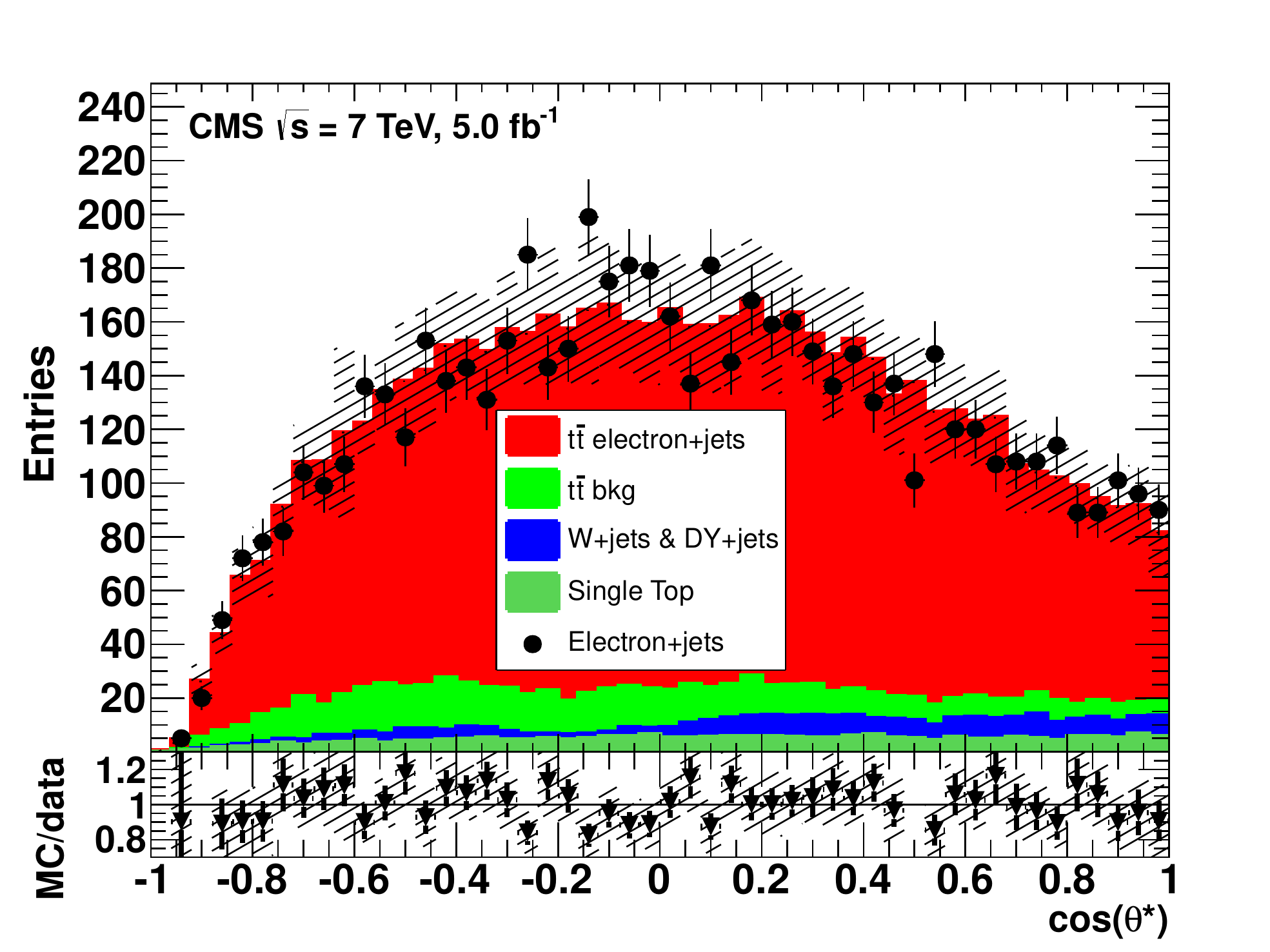} }

\centerline{ \includegraphics[width = 0.45\textwidth]{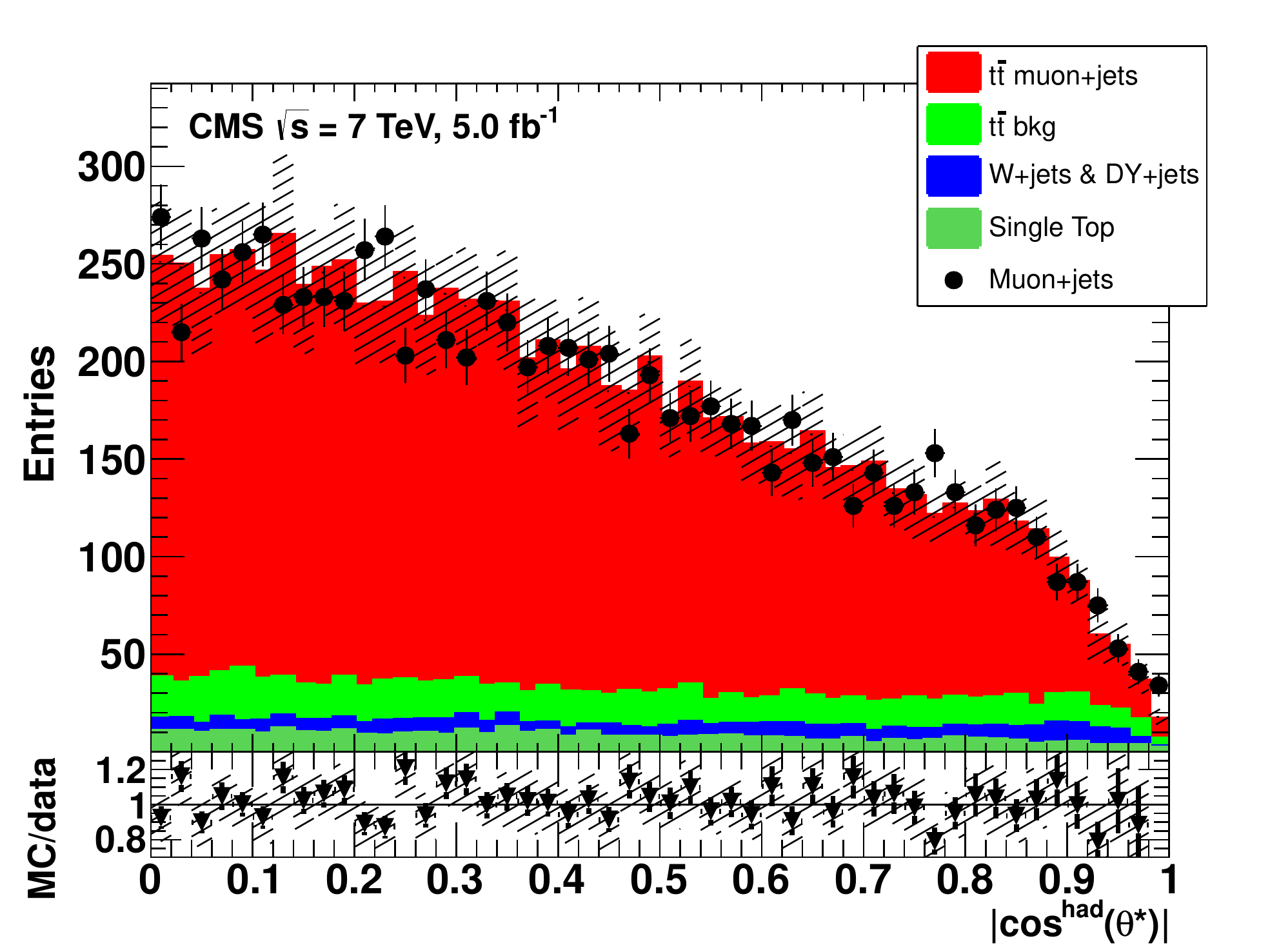} \includegraphics[width = 0.45\textwidth]{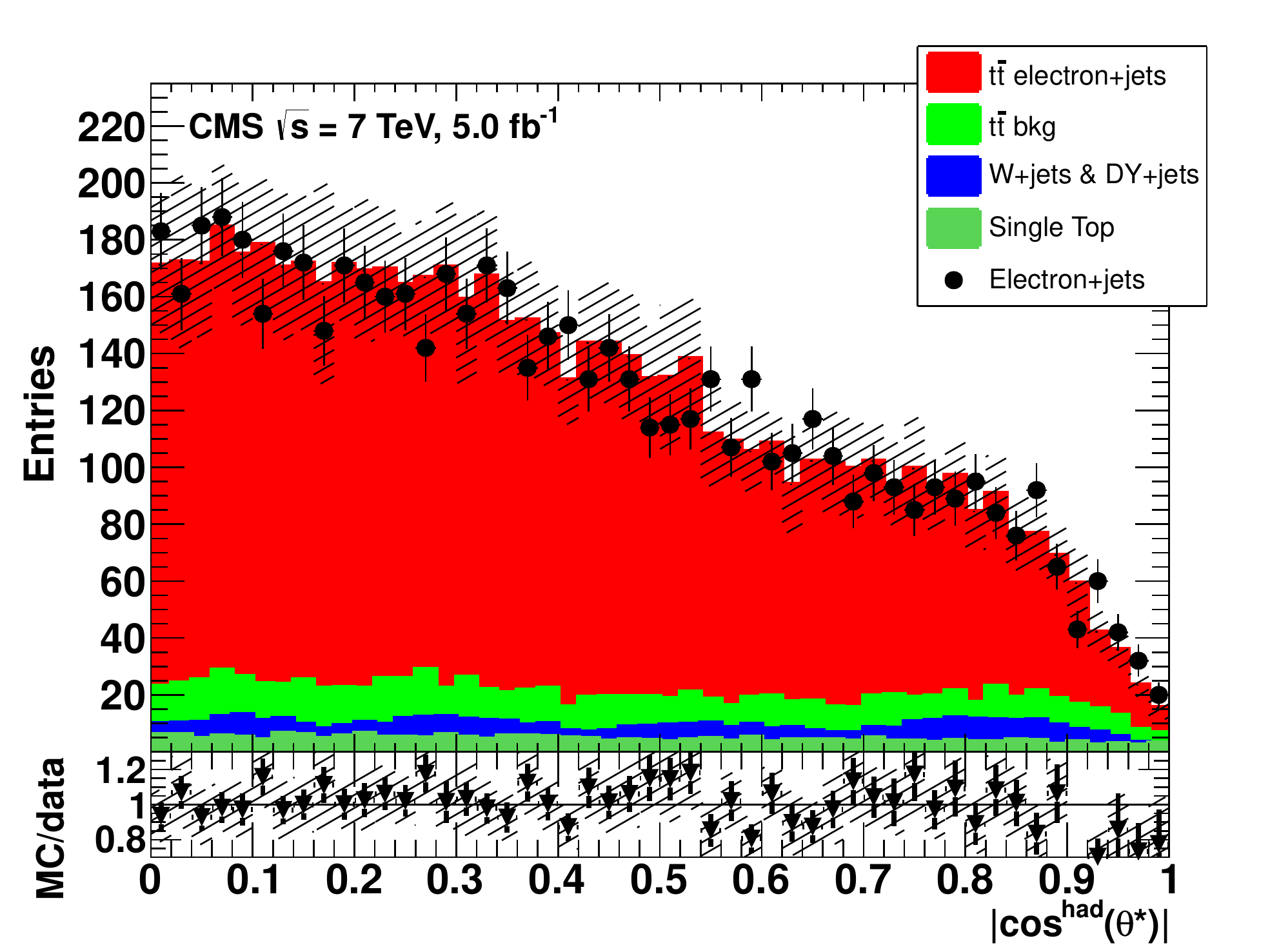} }
\caption{
Cosine of the helicity angles \costh~(top) and $\abs{\costhhad}$ (bottom) for the
muon+jets (left) and electron+jets (right) channels. Data are displayed as
solid points, simulated \ttbar signal distributions as red histograms, and the
contribution from other background processes as coloured histograms.
At the bottom,
the ratio between prediction and data is displayed.  Systematic
uncertainties are shown as hatched histograms.
 \label{fig:costh}}
\end{figure}

A comparison of shapes between data and simulation, after having applied the kinematic
fit, for the variables
relevant to this analysis is presented in figures~\ref{fig:controlmu} to
\ref{fig:costh}.  Figure~\ref{fig:controlmu} shows the
kinematic distributions for the leptons in the event: transverse momentum and pseudorapidity of muons and electrons,
and the neutrino transverse momentum.
Figure~\ref{fig:jetNu} displays the \pt distributions for
jets related to the top-quark reconstruction, including the
b jets (from leptonic and hadronic tops)
and the
jets from the hadronic W-boson decay.
The shapes of all important kinematic distributions in the data, which
describe the \ttbar system, are well reproduced by the simulation,
including those that do not depend strongly on the W-boson helicity (\ie
global properties of top-quark and W-boson systems).

The most relevant distributions for this analysis, the cosine distributions of
the helicity angles computed according to the definitions discussed in section
\ref{sec:intro}, are shown in figure~\ref{fig:costh} for both the
muon+jets and electron+jets channels.
The agreement
between data and simulation for \costh ~and $\abs{\costhhad}$ is observed to
be  satisfactory.  This suggests that, even before any attempt to
measure the W-boson helicity fractions is made, the data prefer values close to
the SM predictions.

\section{Systematic uncertainties}
\label{sec:syst}

Several sources of systematic effects, which could possibly bias the
measurement of the W-boson helicity fractions, have been investigated and their
corresponding uncertainties in the measurement determined.

The scale of the jet energy (JES) calibration is determined from data,
which is then applied as a \pt- and $\eta$-dependent correction to
the simulation; the JES calibration has an uncertainty that
typically varies between 2\% and 4\%~\cite{CMSjes}.  To
estimate the effect that the JES uncertainty has on the W-boson helicity
measurement, the \pt of all jets are systematically shifted
together, either up or down, by their corresponding \pt- and
$\eta$-dependent uncertainty. Because the missing transverse energy is
corrected due to the presence of JES calibrated jets, the systematic shifts
of the jet \pt's in the event are propagated and lead to systematic
shifts in the
momentum imbalance.
The full analysis, including
the reconstruction of top quarks and the resulting measurements of the W-boson
helicity fractions, is then repeated.
The jet energy resolution (JER) is observed to be different in data
compared with the simulation. Jet energy resolutions
are typically 5--13\% larger in data than in simulation, with uncertainties smaller
than 5\%~\cite{CMSjes}.
The systematic uncertainties related to JER are estimated by over-smearing the reconstructed jets in
simulated events, so that their transverse momentum resolution is the same as
measured in the data. The effect is propagated to the missing transverse energy,
and the full analysis is repeated, similar to the procedure used to estimate the JES uncertainty.

Uncertainties in the lepton identification efficiency are investigated by
varying the efficiency correction factor
$\epsilon_{\ell}^\text{DATA}/\epsilon_{\ell}^\mathrm{MC}$.
In the case of muons, the efficiency correction factor
depends on the $\eta$ position of the muon in the detector.
Since the measurement of the W-boson helicity is mainly affected by shape-dependent effects,
uncertainties due to the muon efficiency correction factor are estimated by repeating
the full analysis replacing the $\eta$-dependent factors by an uniform correction.
 In the case of
electrons, only a very mild dependency on $\eta$ is observed and the
corresponding uncertainty, derived by assuming a flat $\eta$-dependence,
has very little impact on the W-boson helicity measurement.
Therefore,
the efficiency correction factor for electron identification is shifted together
with a shift in the scale factor for the jet component of the trigger
according to the \pt and $\eta$ position of the electrons and jets.
The combination of electron and jet scale factors that lead to the maximum
possible $\eta$-dependent effect is then applied and the full analysis is repeated.

The efficiencies for
b tagging are measured using a control sample of
multijet events~\cite{ref:btag}, in data and simulation. The correction factors
${\epsilon_\text{\cPqb-tag}^\text{DATA}}/{\epsilon_\text{\cPqb-tag}^\mathrm{MC}}$, which are applied to
the simulated samples, are functions of the \pt and $\eta$ of the jets, the number of
the required b tags, as well as the number of heavy-flavour and light-flavour jets in the event. The scale
factors are relatively uniform in the typical \pt range of the selected
jets, resulting in only a very small dependence of the W-boson helicity results on the b-tagging
efficiency. The scale factors are varied by their uncertainties and the resulting
differences are taken as a systematic uncertainty in the measured fractions.

The effect of pileup is estimated by varying the
number of minimum bias collisions superimposed on each simulated signal
event, according to an uncertainty of 5\% which includes the
uncertainties in the inelastic pp cross section, luminosity and other modelling uncertainties.

To account for a bias on the W-boson helicity measurement due to
uncertainties in the normalisation from simulated background samples
involving single top quarks, relative to the signal, the assumed reference
cross sections are varied.  While CMS has measured the single-top-quark
production cross section in the $t$-channel with a 9\% precision ($67.2\pm
6.1$\unit{pb}~\cite{CMSsingletchan}), a systematic variation of
$\pm$15\% is applied to cover the case of single-top-quark events produced with
additional jets from radiation, which comprise the main contribution to
this background component. Likewise, while CMS measures a cross section of
16$^{+5}_{-4}$\unit{pb}~\cite{CMSsingletWchan} for the associated tW production
case, the reference cross section used for the normalisation, 15.74\unit{pb}, is
shifted by $\pm$40\%.  Finally, since the \ttbar sample
normalisation is a free parameter in the helicity fits, the uncertainty
associated with its initially assumed reference cross section does not
affect the measurement.

The normalisation of the  W+jets and DY+jets samples is estimated using the method described in section \ref{sec:bkg}
with an uncertainty of 100\% and 30\%, respectively.
In both the W+jets and DY+jets events
cases, the shapes of all relevant distributions in the simulation agree
with the data and any systematic effects from variations in normalisations
are much larger than any variations arising from shape; hence, systematic
uncertainties due to possible differences in shape are negligible.

While the sample size for the reference simulated \ttbar dataset is chosen to be five times that of the processed data,
it
is possible for the
reweighting method to introduce a systematic bias by degrading the
statistical power of the simulated sample due to weights which can
be larger than unity.  Hence, special care must be taken to ensure that the
statistical uncertainty of the MC prediction for each
$\cos\theta^*_\text{rec}$ bin is substantially smaller than the corresponding
statistical uncertainty of the data.
The uncertainties are estimated by repeating the analysis using a subsample of events that correspond to a fraction $1/N$
of the entire sample. The procedure is repeated many times, and the uncertainties
on the W-boson helicity fractions taken as $\sigma/\sqrt{N}$ where $\sigma$ is the spread observed on the fraction.
Several values of $N$, between 2 and 10, are tested, and result in very similar uncertainties.

Since the W-boson helicity fractions depend directly on the top-quark mass,
uncertainties in the latter 
could bias the measurement.  This
systematic effect is studied and taken into account, via \ttbar samples
simulated using \MADGRAPH for different $m_\cPqt$ hypotheses.  A
variation of $\pm$1.4\GeVcc~\cite{topmass} about the assumed central value of 172.5\GeVcc
is assumed.

Uncertainties on the helicity measurement from the choice of
renormalisation and factorisation scales
 for the simulated signal samples are estimated using dedicated \ttbar simulated
samples that vary the renormalisation and factorisation scales  
and the scale of the first emission in
the parton shower, in a consistent manner, by factors of 0.5 and 2
with respect to a central value of $Q$, with $Q^2= m_\cPqt^2c^2+(\sum \pt^\text{jet})^2$. The kinematic scale used to match jets to partons
in the signal simulation is estimated using dedicated \ttbar samples where that matching parameter is varied
by factors of 0.75 and 1.5 with respect to its central value of 40\GeV.

The systematic effects due to the PDFs
used to simulate the signal and background
samples are estimated using two different methods~\cite{pdf4lhc}, according to a
reweighting technique.  Firstly, events are reweighted using 100 members~\cite{lhapdf}
of the NNPDF21 set~\cite{NNPDF}, and the W-boson helicity fractions are
remeasured for each of them. For a given helicity fraction, the RMS of
the distribution of measurements from the
different
PDF members
provides an uncertainty estimate that corresponds to 68\% confidence level~(CL).
Secondly, the difference between the central values for CTEQ6L1~\cite{cteq}  (used in the analysis simulations) and MSTW2008lo68cl~\cite{pdfs} is estimated. The systematic uncertainties in the
measurements for the W-boson helicity fractions, due to intrinsic PDF
uncertainties, are then taken as the largest difference between two
different estimates.

The impact of all of the above systematic effects on the W-boson helicity fractions is
detailed in table~\ref{tab:system2blep}, for the measurements for the
leptonic side of the events (\costh). The table shows results for the 3D
and 2D fits, obtained by fitting two of the fractions or setting $F_R=0$,
respectively.  Measurements using final states containing either a muon or
an electron are presented in columns two through seven. The last three
columns display systematic uncertainties for the combined measurements of
the muon+jets and electron+jets channels, using the measurements from
columns two through seven as inputs.  The measurements are combined taking
into account both statistical and systematic uncertainties. Common sources
of uncertainties between the different measurements are assumed to be fully
correlated. The dominant sources of systematic uncertainties in the
leptonic side are the W+jets background normalisation, the signal
modelling (\ttbar renormalisation and factorization scales, and top-quark mass), and
the statistics of the simulated samples.

Systematic uncertainties in the measurements for the hadronic branch, using
the $\abs{\costhhad}$, are presented in table \ref{tab:systemHad}.  Since
$\abs{\costhhad}$ has no sensitivity to a measurement of $F_L-F_R$, only the
2D fits are performed.  Uncertainties on the individual measurements in the
electron and muon channels are presented in the first two columns; in the
last column, the combination of muons+jets and electrons+jets channels is
shown.  The hadronic branch is seen to have larger systematic
uncertainties, compared with the leptonic branch, due, in part, to the
dominant W+jets background and the importance of uncertainties from the JES
and JER, as well as PDFs.

\begin{table}[th]
\begin{center}
\topcaption{Summary of the systematic uncertainties for the analysis using only the leptonic branch of the event, for the 3D fit,
  fitting   $F_0, F_L$, and
   $\mathcal{F}_{\ttbar}$ (columns 2--3 for muon+jets analysis, 5--6 for electron+jets analysis, and 8--9 for the combination
of both decay modes); and  the  2D fit, fitting $F_0$ and
   $\mathcal{F}_{\ttbar}$  only  (column 4 for muon+jets analysis, 7 for electron+jets analysis, and 10 for the combination of both
decay modes). The numbers given
  correspond to the absolute uncertainty with respect to  the central analysis:
$\Delta F=(F^\text{central}-F^\text{check})$.
\label{tab:system2blep}}
\begin{tabular}{|l|c|c||c||c|c||c||c|c||c|} \hline
 &  \multicolumn{3}{c||}{$\mu$+jets (\costh)}  &  \multicolumn{3}{c ||}{e+jets (\costh)}
 &  \multicolumn{3}{c |}{$\ell$+jets (\costh)}  \\  \cline{2-10}
 Systematic  & \multicolumn{2}{c||}{  { 3D fit} }  &  { 2D fit}
  & \multicolumn{2}{c||}{  { 3D fit} }  &  { 2D fit}  & \multicolumn{2}{c||}{  { 3D fit} }  &  { 2D fit}   \\ \cline{2-10}
 Uncertainties & $\pm$  $\Delta F_0$  &  $\pm$ $\Delta F_L$  &  $\pm$  $\Delta F_0$  & $\pm$ $\Delta F_0$  &  $\pm$ $\Delta F_L$  &  $\pm$  $\Delta F_0$    & $\pm$  $\Delta F_0$  &  $\pm$ $\Delta F_L$  &  $\pm$  $\Delta F_0$    \\
\hline \hline
         JES        &  0.005 & 0.003  &   0.001  &  0.006 & 0.002 &  0.003  & 0.006 & 0.003 & 0.001 \\
          JER       &  0.009 & 0.005  &   0.001  &  0.014 & 0.009 &  0.003  & 0.011 & 0.007 & 0.001 \\
        Lepton eff. &  0.001 & 0.001  &   0.001  &  0.009 & 0.012 &  0.015  & 0.001 & 0.002 & 0.002  \\
      b-tag eff.    &  0.001 & 0.001  &   {\small $<10^{-3}$}  &  {\small $<10^{-3}$} & {\small $<10^{-3}$} &  0.001  & 0.001 & {\small $<10^{-3}$} & {\small $<10^{-3}$} \\
           Pileup   &  0.013 & 0.011  &   0.008  &  0.008 & 0.007 &  0.005  & 0.002 & {\small $<10^{-3}$} & 0.008 \\
      Single-t bkg. &  0.004 & {\small $<10^{-3}$}  &   0.003  &  0.004 & {\small $<10^{-3}$} &  0.004  & 0.004 & 0.001 & 0.003 \\
        W+jets bkg. &  0.019 & 0.007  &   0.006  &  0.009 & 0.006 &  0.022  & 0.013 & 0.004 & 0.006 \\
       DY+jets bkg. &  0.002 & 0.001  &   0.001  &  0.001 & {\small $<10^{-3}$} &  0.001  & 0.001 & {\small $<10^{-3}$} & 0.001 \\
     MC statistics  &  0.016 & 0.012  &   0.009  &  0.019 & 0.015 &  0.012  & 0.016 & 0.012 & 0.010 \\
      Top-quark mass&  0.011 & 0.008  &   0.007  &  0.025 & 0.018 &  0.014  & 0.016 & 0.011 & 0.019 \\
 \ttbar  scales     &  0.013 & 0.009  &   0.007  &  0.015 & 0.018 &  0.030  & 0.009 & 0.009 & 0.011 \\
 \ttbar match. scale&  0.004 & 0.001  &   0.006  &  0.010 & 0.013 &  0.016  & 0.011 & 0.010 & 0.008 \\
            PDF     &  0.002 & 0.001  &   0.003  &  0.004 & 0.002 &  0.002  & 0.002 & {\small $<10^{-3}$} & 0.003 \\
\hline
\end{tabular}
\end{center}
\end{table}

\begin{table}[ht]
\begin{center}
\topcaption{Systematic uncertainties for the 2D fits using the hadronic branch of the \ttbar system, and for the muon channel, electron channel, as well as the combination of both decay channels.  The numbers given
  correspond to the absolute uncertainty with respect to  the central analysis:
$\Delta F=(F^\text{central}-F^\text{check})$. \label{tab:systemHad} }
\begin{tabular}{|l|c|c||c|} \hline
    & $\mu$+jets ($\abs{\costhhad}$) & e+jets ($\abs{\costhhad}$)  &    $\ell$+jets ($\abs{\costhhad}$)  \\ \cline{2-4} 
  Systematic   & 2D fit & 2D fit & 2D fit  \\ \cline{2-4}
  Uncertainties & $\pm$  $\Delta F_0$   & $\pm$  $\Delta F_0$   & $\pm$  $\Delta F_0$  \\ \hline \hline
         JES        &  0.010 &  0.008 & 0.002  \\
          JER       &  0.042 &  0.032 & 0.038  \\
        Lepton eff. &  0.002 &  0.002 &  0.001 \\
      b-tag eff.    &  0.003 &   {\small $<10^{-3}$} &  0.002 \\
           Pileup   &  0.018 &  0.006 & 0.015  \\
      Single-t bkg. &  0.005 &  0.007 &  0.006 \\
        W+jets bkg. &  0.060 &  0.050 &  0.040 \\
       DY+jets bkg. &  0.002 &  0.005 &  0.002 \\
     MC statistics  &  0.023 &  0.028 & 0.025 \\
      Top-quark mass&  0.008 &  0.041 & 0.014  \\
 \ttbar  scales     &  0.022 &  0.033 & 0.027  \\
 \ttbar match. scale&  0.002 &  0.035 & 0.013 \\
            PDF     &  0.013 &  0.014 & 0.014  \\
\hline
\end{tabular}
\end{center}
\end{table}

\section{Results}
\label{sec:results}

The W-boson helicity fractions are measured according to the fits described in
section \ref{sec:fit}.
The unitary condition $F_0+F_L+F_R=1$ is used to determine either (a) the
right-handed fraction $F_R$ from measurements of the free parameters, $F_0$
and $F_L$, in the 3D fits or (b) the left-handed fraction $F_L$ from the
measurement of the free parameter $F_0$ in the 2D fits assuming $F_R=0$.
Table \ref{table:individual} presents the fit measurement of each helicity
parameter, one decay channel at a time, together with the statistical and
systematic uncertainties.
The statistical correlation factor $\rho_{0L}^\text{stat}$  between $F_0$ and $F_L$ is presented in
the last column; the measurements are seen to be highly
correlated.
All measurements, from either the muon or electron channels, using either
the leptonic or hadronic branches, are observed to be compatible within
uncertainties.
The measurements using the leptonic branch \costh ~are more precise, as
expected.

\begin{table}[htp]
\begin{center}
\topcaption{Measurements of the W-boson helicity fractions from the \costh
 (leptonic branch) and $\abs{\costhhad}$ (hadronic branch) distributions. The
columns show the fit type, the decay channel, and the measurement of each
helicity parameter, together with the statistical and systematic
uncertainties. For the 3D fits, the last column presents the statistical
correlation between $F_0$ and $F_L$, while for the 2D fit, total anticorrelation ($F_L=1-F_0$) is assumed.
\label{table:individual}}
{\small
\begin{tabular}{|l|l||r|r|r||c|} \hline
\multicolumn{6}{|c|}{Leptonic branch: \costh } \\ \hline\hline
Fit &  Channel &  $F_0$ $\pm$ (stat.) $\pm$ (syst.) & $F_L$ $\pm$ (stat.)
$\pm$ (syst.)  & $F_R$ $\pm$ (stat.) $\pm$ (syst.) & $\rho_{0L}^\text{stat}$
\\ \hline
3D  & $\mu$+jets    &   0.674 $\pm$0.039$\pm$0.035    & 0.314
$\pm$0.028$\pm$0.022    &    0.012 $\pm$0.016$\pm$0.020  & $-0.95$ \\
3D  & e+jets        &   0.688 $\pm$0.045$\pm$0.042    & 0.310
$\pm$0.033$\pm$0.037    &    0.002 $\pm$0.017$\pm$0.023  & $-0.95$ \\ \hline
2D  & $\mu$+jets    &   0.698 $\pm$0.021$\pm$0.019    & 0.302
$\pm$0.021$\pm$0.019    &    fixed at 0 ~ ~  ~ ~  ~ ~ ~  & $-1$ \\
2D  & e+jets        &   0.691 $\pm$0.025$\pm$0.047    & 0.309
$\pm$0.025$\pm$0.047    &    fixed at 0 ~ ~  ~ ~  ~ ~ ~  & $-1$ \\ \hline\hline
\multicolumn{6}{|c|}{Hadronic branch: $\abs{\costhhad}$ } \\ \hline\hline
Fit &  Channel &  $F_0$ $\pm$ (stat.) $\pm$ (syst.) & $F_L$ $\pm$ (stat.)
$\pm$ (syst.)  & $F_R$ $\pm$ (stat.) $\pm$ (syst.) & $\rho_{0L}$
\\ \hline
2D  & $\mu$+jets      &   0.651 $\pm$0.060$\pm$0.084  & 0.349
$\pm$0.060$\pm$0.084    &    fixed at 0 ~ ~  ~ ~  ~ ~ ~  & $-1$ \\
2D  & e+jets          &   0.629 $\pm$0.060$\pm$0.093  & 0.371
$\pm$0.060$\pm$0.093    &    fixed at 0 ~ ~  ~ ~  ~ ~ ~  & $-1$ \\ \hline
\end{tabular}
}
\end{center}
\end{table}

Table \ref{table:combined} presents various combinations of the results
presented in table \ref{table:individual}.
Firstly, the muon+jets and electron+jets channels are combined using the
leptonic branch measurements from the 3D fits.
The $\chi^2$ per degree of freedom for that combination is 0.109/2,
corresponding to a $\chi^2$-probability of 94.7\%.
Secondly, the 2D fit measurements of the $F_0$ helicity fraction from the
leptonic (\costh) ~and hadronic (\costhhad) ~branches are combined,
separately for each decay channel.
While the leptonic branch dominates with a weight of about 90\%, the total
uncertainty of the combination nevertheless decreases.
Finally, the most precise measurement of $F_0$ is obtained by subsequently
combining the 2D fit measurements across the muon+jets and electron+jets
channels, following the combination of the 2D fit measurements from the
leptonic and hadronic branches.

\begin{table}[htp]
\begin{center}
\topcaption{The combined helicity fractions and their uncertainties, including
the type of fit performed, the channels ($\ell={\rm e},\mu$ combination) and branches of the \ttbar system
("l" for leptonic, \costh, and "h" for hadronic, $\abs{\costhhad}$, used in the
combination, as well as the total correlation between $F_0$ and $F_L$.
\label{table:combined} }
\begin{tabular}{|c|c|c|l|r|c|}\hline
Fit &  Channel(s)  & Branch & \multicolumn{2}{|c|}{Fraction $\pm$ (stat.)
$\pm$ (syst.) [total] }  & $\rho_{0L}^\text{total}$ \\ \hline\hline
    &              &        & $F_0$ & 0.682 $\pm$0.030$\pm$0.033 [0.045] &
       \\
3D  & $\ell$+jets      & l      & $F_L$ & 0.310 $\pm$0.022$\pm$0.022 [0.032] &
$-0.95$  \\
    &              &        & $F_R$ & 0.008 $\pm$0.012$\pm$0.014 [0.018] &
       \\ \hline
2D  & $\mu$+jets        & l+h    & $F_0$ & 0.694 $\pm$0.020$\pm$0.025 [0.032] &
       \\
    &              &        & $F_L$ & 0.306 $\pm$0.020$\pm$0.025 [0.032] &
$-1$     \\ \hline
2D  & e+jets            & l+h    & $F_0$ & 0.674 $\pm$0.025$\pm$0.028 [0.037] &
       \\
    &              &        & $F_L$ & 0.326 $\pm$0.025$\pm$0.028 [0.037] &
$-1$     \\ \hline
2D  & $\ell$+jets      & l+h    & $F_0$ & 0.685 $\pm$0.017$\pm$0.021 [0.027] &
       \\
    &              &        & $F_L$ & 0.315 $\pm$0.017$\pm$0.021 [0.027] &
$-1$     \\ \hline
\end{tabular}
\end{center}
\end{table}

\begin{figure}[hp]
\centering
\includegraphics[scale=0.8]{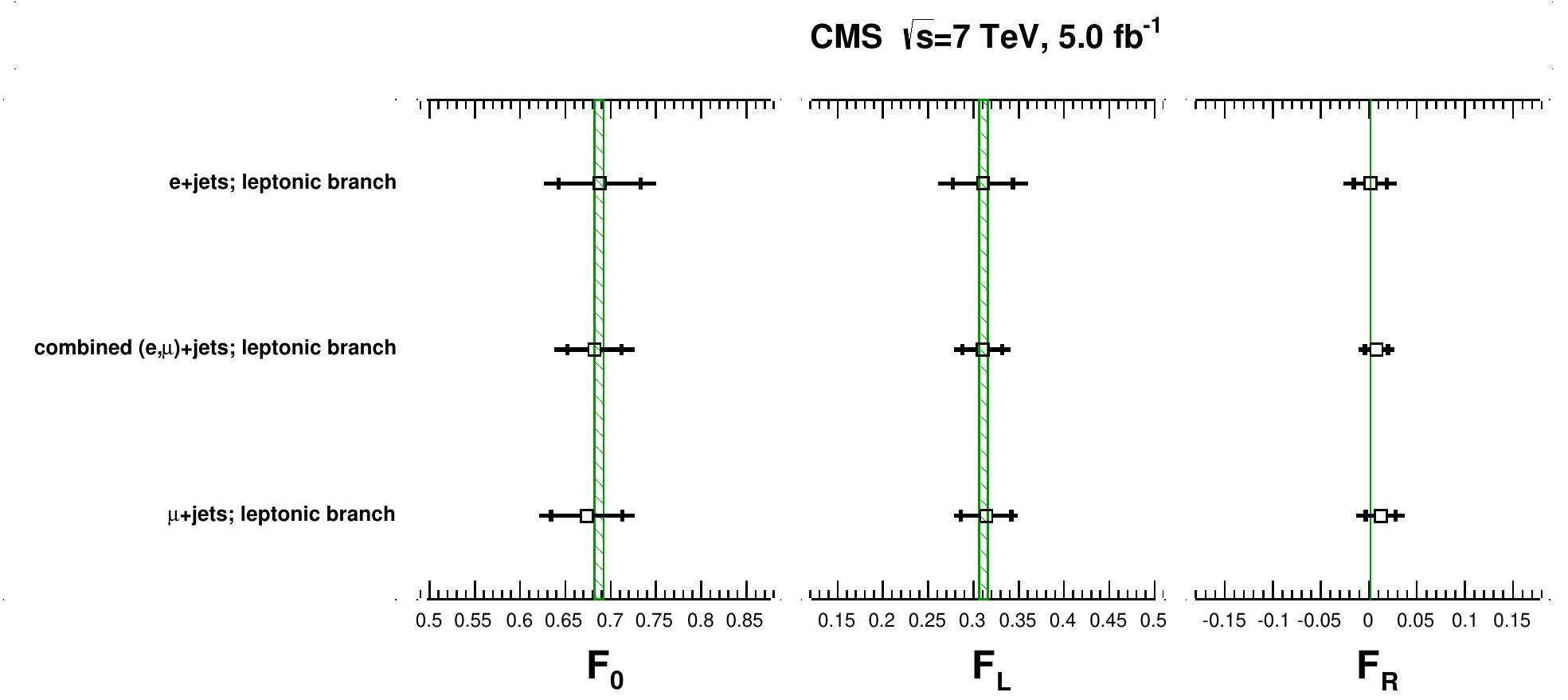}
\caption{Summary of the W-boson helicity measurements in semileptonic decays of
top-quark pairs with 2011 data for 3D fits. The inner error bars represent the statistical uncertainties
and the outer error bars the statistical and systematic uncertainties, added in quadrature.
 NNLO predictions from ref.~\cite{Czarnecki:2010gb}
with their theoretical uncertainties are represented as hatched
bands. \label{money1} }

\end{figure}

\begin{figure}[hp]
\centering
\includegraphics[scale=0.8]{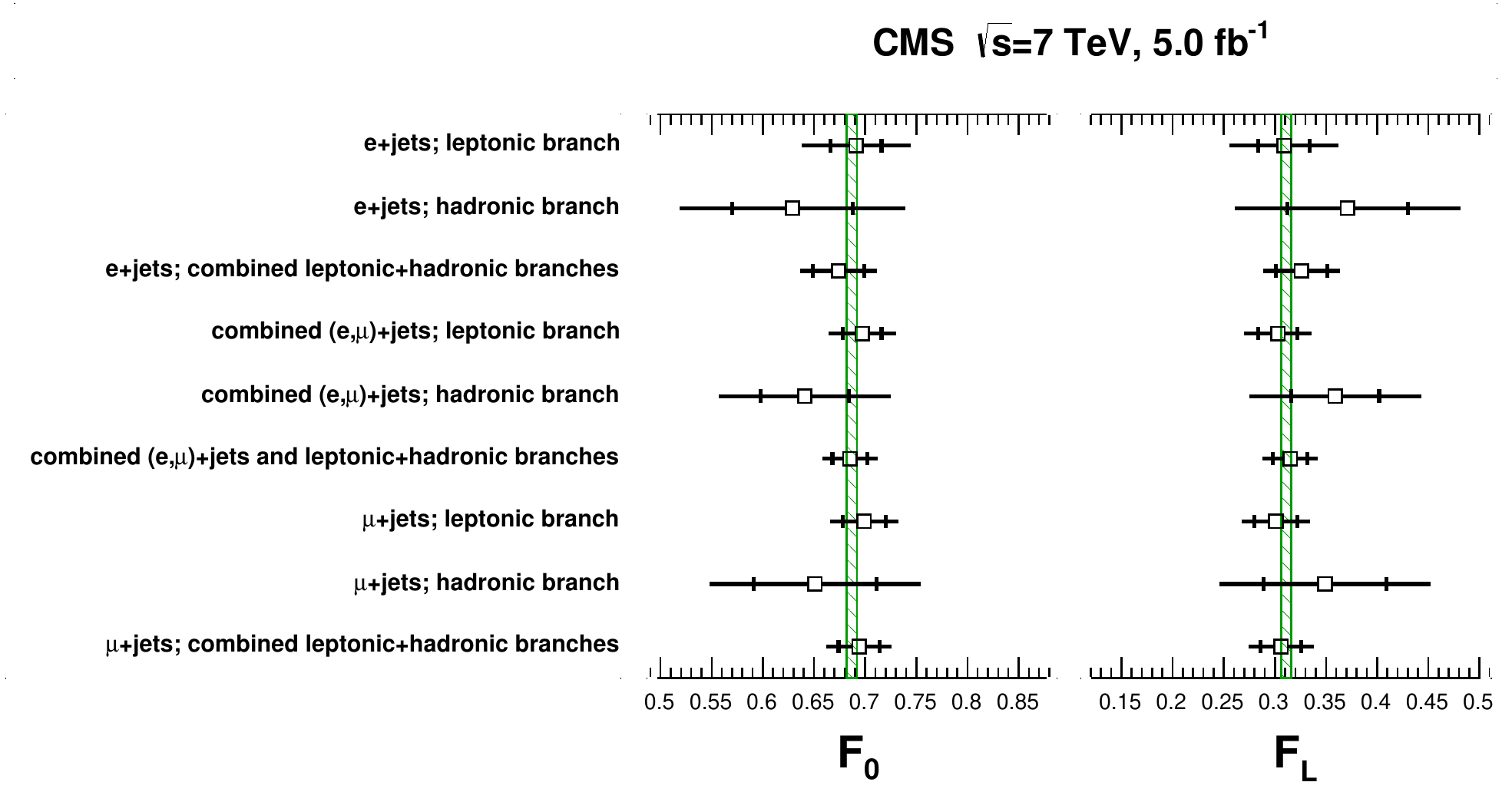}
\caption{Summary of the W-boson helicity measurements in semileptonic decays of
top-quark pairs with 2011 data, for 2D fits assuming $F_R=0$.
The inner error bars represent the statistical uncertainties
and the outer error bars the statistical and systematic uncertainties, added in quadrature.
NNLO predictions from ref.~\cite{Czarnecki:2010gb}
with their theoretical uncertainties are represented as hatched
bands. \label{money2} }
\end{figure}

Summaries of all measurements and their various combinations are presented
in figures \ref{money1} and \ref{money2} for the 3D and 2D types of fits,
respectively.
All measurements are compatible with each other, and also compatible with
the expectations from the SM~\cite{Czarnecki:2010gb}.

\section{Limits on anomalous couplings}

The measured  helicity fractions can be used to set limits on anomalous Wtb
couplings.
We assume the minimal parametrisation of the Wtb vertex suggested in
refs.~\cite{eqSaav1,eqSaav2,eqSaav3} and as described in the introduction.
We consider two specific scenarios.
First, we assume $V_L=1, V_R=g_L=0$ and leave $\Re(g_R)$ as a free
parameter.
This CP-conserving scenario is particularly interesting because indirect
constraints to $g_R$ from radiative B-meson decay measurements are
currently poor, $\Re(gR)\in [-0.15,+0.57]$~\cite{AnomBphysics}.
A specific feature of this scenario is that it does not provide any
contribution to the right-handed helicity of the W boson, $F_R$.
The grand combination of the longitudinal helicity fraction $F_0$
measurements, across both the leptonic and hadronic branches including both
the muon and electron channels, and assuming $F_R=0$,
is reinterpreted in terms of $\Re(g_R)$, yielding
\begin{equation*}
\Re(g_R) = -0.008 \pm 0.024\stat^{+0.029}_{-0.030}\syst,
\end{equation*}
\noindent
which is consistent with the SM expectations within the quoted
uncertainties.
In quoting this result we have omitted another minimum of the fit closer to the
$\Re(g_R)\approx 0.8$ region, which would lead to an increase of
almost a factor of three in the single-top-quark cross
section~\cite{SaavedraSingletop},
which would not be consistent
with the recent CMS measurement~\cite{CMSsingletchan}.
In terms of the effective dimension-six Lagrangian $\mathcal{O}^{33}_{uW}$
defined in refs.~\cite{eqSaav1,eqSaav2} we obtain the equivalent result,
\begin{equation*}
 \Re(C^{33}_{uW})/\Lambda^2 = -0.088 \pm 0.280\stat^{+0.339}_{-0.352}\syst\TeV^{-2},
\end{equation*}
where $\Lambda$ is the scale of new physics and $\Re(C^{33}_{uW})$ the
effective operator coefficient.

\begin{figure}[ht]
\centering
\includegraphics[scale=0.6]{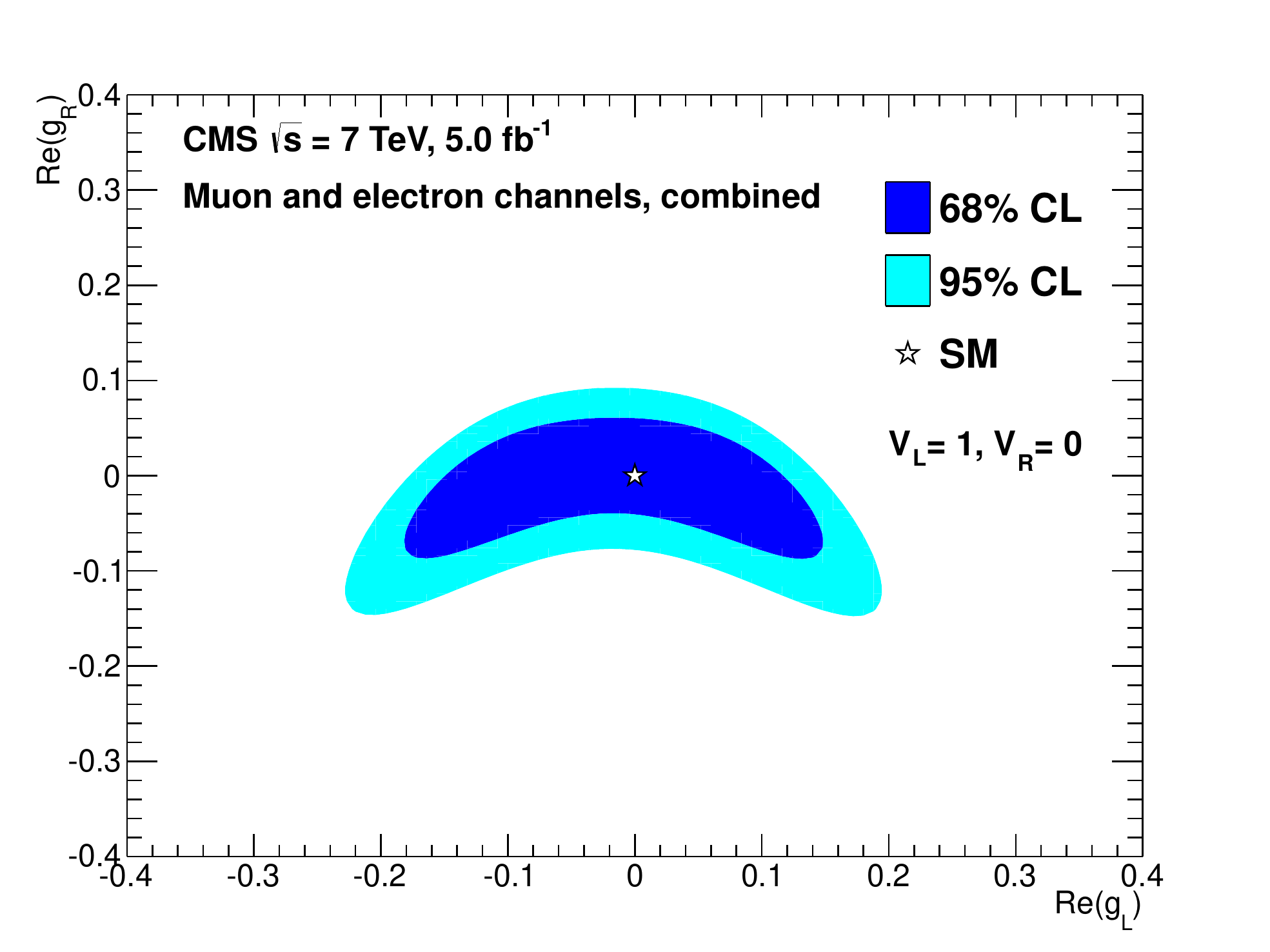}
\caption{ Limits on the real components of the anomalous couplings $g_L,$\
$g_R$ at 68\% and 95\% CL, for $V_L=1$ and $V_R=0$. The SM prediction
($g_R=0$ and $g_L=0$) is also shown.
  \label{fig:limanom}}

\end{figure}

In the second scenario, again assuming CP is conserved, we choose $\Re(g_L)$ and $\Re(g_R)$ as free parameters of the fit.
Limits on those parameters are determined using the combined measurements of the
muon+jets and electron+jets channels from the 3D fit of the leptonic
branch, \costh.
The results of the likelihood
fit for the parameters $F_0$ and $F_L$ can be reinterpreted in terms of the parameters
$\Re(g_L)$ and $\Re(g_R)$.
Figure \ref{fig:limanom} shows the regions of
the $\Re(g_L),\ \Re(g_R)$ plane allowed at 68\% and 95\% CL.
As in the first scenario, a region near $\Re(g_L)=0$ and $\Re(g_R) \gg 0$, allowed by the fit but excluded by the CMS single-top
quark measurement, is not shown.

The result obtained from the first scenario represents an improvement of
about 50\% on the precision of $\Re(g_R)$ with respect to previous
measurements~\cite{AtlasPaper},
while the limits from the second scenario are similar to those from ref.~\cite{AtlasPaper}.

\section{Summary}

The W-boson helicity has been measured in top-quark-pair events decaying
semileptonically, both in the  muon+jets and in the electron+jets channels,
using proton-proton collisions data at $\sqrt{s}=7$\TeV, corresponding to an integrated luminosity of 5.0\fbinv.
Both the leptonic and the hadronic branches of the decay have been studied.
The most precise measurement, not constraining $F_R$ to the SM,  corresponding to the combination of muon and
electron channels, using only the leptonic branch,  yields:
\begin{align*}
F_0 &= 0.682 \pm  0.030\stat\pm 0.033\syst, \\
F_L &= 0.310 \pm  0.022\stat\pm 0.022\syst, \\
F_R &=  0.008 \pm 0.012\stat\pm 0.014\syst,
\end{align*}
 with a correlation coefficient of $-$0.95 between $F_0$ and $F_L$.

The measured W-boson helicity fractions are in agreement with the predictions
from the standard model.
Assuming a minimal parametrisation of the Wtb vertex, stringent limits on
the real components of the anomalous couplings $g_L$ and $g_R$ are also derived.

\section*{Acknowledgements}

\hyphenation{Bundes-ministerium Forschungs-gemeinschaft Forschungs-zentren} We congratulate our colleagues in the CERN accelerator departments for the excellent performance of the LHC and thank the technical and administrative staffs at CERN and at other CMS institutes for their contributions to the success of the CMS effort. In addition, we gratefully acknowledge the computing centres and personnel of the Worldwide LHC Computing Grid for delivering so effectively the computing infrastructure essential to our analyses. Finally, we acknowledge the enduring support for the construction and operation of the LHC and the CMS detector provided by the following funding agencies: the Austrian Federal Ministry of Science and Research and the Austrian Science Fund; the Belgian Fonds de la Recherche Scientifique, and Fonds voor Wetenschappelijk Onderzoek; the Brazilian Funding Agencies (CNPq, CAPES, FAPERJ, and FAPESP); the Bulgarian Ministry of Education and Science; CERN; the Chinese Academy of Sciences, Ministry of Science and Technology, and National Natural Science Foundation of China; the Colombian Funding Agency (COLCIENCIAS); the Croatian Ministry of Science, Education and Sport; the Research Promotion Foundation, Cyprus; the Ministry of Education and Research, Recurrent financing contract SF0690030s09 and European Regional Development Fund, Estonia; the Academy of Finland, Finnish Ministry of Education and Culture, and Helsinki Institute of Physics; the Institut National de Physique Nucl\'eaire et de Physique des Particules~/~CNRS, and Commissariat \`a l'\'Energie Atomique et aux \'Energies Alternatives~/~CEA, France; the Bundesministerium f\"ur Bildung und Forschung, Deutsche Forschungsgemeinschaft, and Helmholtz-Gemeinschaft Deutscher Forschungszentren, Germany; the General Secretariat for Research and Technology, Greece; the National Scientific Research Foundation, and National Office for Research and Technology, Hungary; the Department of Atomic Energy and the Department of Science and Technology, India; the Institute for Studies in Theoretical Physics and Mathematics, Iran; the Science Foundation, Ireland; the Istituto Nazionale di Fisica Nucleare, Italy; the Korean Ministry of Education, Science and Technology and the World Class University program of NRF, Republic of Korea; the Lithuanian Academy of Sciences; the Mexican Funding Agencies (CINVESTAV, CONACYT, SEP, and UASLP-FAI); the Ministry of Science and Innovation, New Zealand; the Pakistan Atomic Energy Commission; the Ministry of Science and Higher Education and the National Science Centre, Poland; the Funda\c{c}\~ao para a Ci\^encia e a Tecnologia, Portugal; JINR, Dubna; the Ministry of Education and Science of the Russian Federation, the Federal Agency of Atomic Energy of the Russian Federation, Russian Academy of Sciences, and the Russian Foundation for Basic Research; the Ministry of Education, Science and Technological Development of Serbia; the Secretar\'{\i}a de Estado de Investigaci\'on, Desarrollo e Innovaci\'on and Programa Consolider-Ingenio 2010, Spain; the Swiss Funding Agencies (ETH Board, ETH Zurich, PSI, SNF, UniZH, Canton Zurich, and SER); the National Science Council, Taipei; the Thailand Center of Excellence in Physics, the Institute for the Promotion of Teaching Science and Technology of Thailand, Special Task Force for Activating Research and the National Science and Technology Development Agency of Thailand; the Scientific and Technical Research Council of Turkey, and Turkish Atomic Energy Authority; the Science and Technology Facilities Council, UK; the US Department of Energy, and the US National Science Foundation.
Individuals have received support from the Marie-Curie programme and the European Research Council and EPLANET (European Union); the Leventis Foundation; the A. P. Sloan Foundation; the Alexander von Humboldt Foundation; the Belgian Federal Science Policy Office; the Fonds pour la Formation \`a la Recherche dans l'Industrie et dans l'Agriculture (FRIA-Belgium); the Agentschap voor Innovatie door Wetenschap en Technologie (IWT-Belgium); the Ministry of Education, Youth and Sports (MEYS) of Czech Republic; the Council of Science and Industrial Research, India; the Compagnia di San Paolo (Torino); the HOMING PLUS programme of Foundation for Polish Science, cofinanced by EU, Regional Development Fund; and the Thalis and Aristeia programmes cofinanced by EU-ESF and the Greek NSRF.

\bibliography{auto_generated}   
\cleardoublepage \appendix\section{The CMS Collaboration \label{app:collab}}\begin{sloppypar}\hyphenpenalty=5000\widowpenalty=500\clubpenalty=5000\textbf{Yerevan Physics Institute,  Yerevan,  Armenia}\\*[0pt]
S.~Chatrchyan, V.~Khachatryan, A.M.~Sirunyan, A.~Tumasyan
\vskip\cmsinstskip
\textbf{Institut f\"{u}r Hochenergiephysik der OeAW,  Wien,  Austria}\\*[0pt]
W.~Adam, T.~Bergauer, M.~Dragicevic, J.~Er\"{o}, C.~Fabjan\cmsAuthorMark{1}, M.~Friedl, R.~Fr\"{u}hwirth\cmsAuthorMark{1}, V.M.~Ghete, N.~H\"{o}rmann, J.~Hrubec, M.~Jeitler\cmsAuthorMark{1}, W.~Kiesenhofer, V.~Kn\"{u}nz, M.~Krammer\cmsAuthorMark{1}, I.~Kr\"{a}tschmer, D.~Liko, I.~Mikulec, D.~Rabady\cmsAuthorMark{2}, B.~Rahbaran, C.~Rohringer, H.~Rohringer, R.~Sch\"{o}fbeck, J.~Strauss, A.~Taurok, W.~Treberer-Treberspurg, W.~Waltenberger, C.-E.~Wulz\cmsAuthorMark{1}
% ...
\vskip\cmsinstskip
% ...
1:~~Also at Vienna University of Technology, Vienna, Austria\\
2:~~Also at CERN, European Organization for Nuclear Research, Geneva, Switzerland\\
% ...

\end{sloppypar}
\end{document}